\pdfoutput=1
\documentclass[conference]{IEEEtran}
\IEEEoverridecommandlockouts
\usepackage[noadjust]{cite}
\usepackage{amsmath,amssymb,amsfonts}
\usepackage{algorithmic}
\usepackage{graphicx}
\usepackage{textcomp}
\usepackage{adjustbox}
\usepackage{svg}
\usepackage{tabularray}
\usepackage{xcolor}
\usepackage{pdfpages}
\usepackage{url}
\usepackage{afterpage}

\def\BibTeX{{\rm B\kern-.05em{\sc i\kern-.025em b}\kern-.08em
    T\kern-.1667em\lower.7ex\hbox{E}\kern-.125emX}}

\usepackage{hyperref}
\begin{document}

\title{From Occupations to Tasks: A New Perspective on Automatability Prediction Using BERT\\
}

\author{\IEEEauthorblockN{Dawei Xu}
\IEEEauthorblockA{\textit{University of Technology Sydney} \\
Sydney, Australia \\
dawei.xu@student.uts.edu.au}
\and
\IEEEauthorblockN{Haoran Yang}
\IEEEauthorblockA{\textit{University of Technology Sydney} \\
Sydney, Australia \\
haoran.yang-2@student.uts.edu.au}
\and
\IEEEauthorblockN{Marian-Andrei Rizoiu}
\IEEEauthorblockA{\textit{University of Technology Sydney} \\
Sydney, Australia \\
marian-andrei.rizoiu@uts.edu.au}
\and
\IEEEauthorblockN{Guandong Xu}
\IEEEauthorblockA{\textit{University of Technology Sydney} \\
Sydney, Australia \\
guandong.xu@uts.edu.au}
}

\maketitle

\begin{abstract}
As automation technologies continue to advance at an unprecedented rate, concerns about job displacement and the future of work have become increasingly prevalent. While existing research has primarily focused on the potential impact of automation at the occupation level, there has been a lack of investigation into the automatability of individual tasks. This paper addresses this gap by proposing a BERT-based classifier to predict the automatability of tasks in the forthcoming decade at a granular level leveraging the context and semantics information of tasks. We leverage three public datasets: O*NET Task Statements, ESCO Skills, and Australian Labour Market Insights Tasks, and perform expert annotation. Our BERT-based classifier, fine-tuned on our task statement data, demonstrates superior performance over traditional machine learning models, neural network architectures, and other transformer models. Our findings also indicate that approximately 25.1\% of occupations within the O*NET database are at substantial risk of automation, with a diverse spectrum of automation vulnerability across sectors. This research provides a robust tool for assessing the future impact of automation on the labor market, offering valuable insights for policymakers, workers, and industry leaders in the face of rapid technological advancement.
\end{abstract}

\begin{IEEEkeywords}
Task automatability prediction, Automated Occupation identification, BERT, Transfer learning
\end{IEEEkeywords}

\section{Introduction}
In the era of rapid technological advancement, the fear of large scale unemployment caused by automation technologies has become a significant concern. This concern is not unfounded; a growing body of research suggests that many occupations could be partially or fully automated in the coming years. According to the research conducted by \cite{freyosborne}, over 47\% of current US employment is at high risk of being automated. Another similar study has been performed using job-level data and concluded a less alarming rate of 9\% of employment is at risk \cite{arntz}. However, these studies have largely focused on the potential for automation at the occupation level, leaving a significant gap in our understanding of how individual tasks within these occupations might be automated and multiple studies have shown the effectiveness of assessing the automatability on tasks level instead of occupations \cite{towards, webb_2019_the}. \par
The automatability of tasks is fundamentally a question of understanding the nature and requirements of the task and this information is often embedded in the contextual and semantic information of textual task statements \cite{towards}. For example, the task "Write a report" might be automatable, but "Write a creative story" might not be. And the word ``run" means differently in "run a program" and "run a marathon" \cite{arntz}. Therefore, our research is based on the assumption that the automatability of tasks can be apprehended through the extraction of contextual and semantic information embedded within textual task statements.\par
Due to this circumstance, our research employ the Bidirectional Encoder Representations from Transformers (BERT) model \cite{devlin2018bert} to understand the context and semantics of words in a task statement allowing for a more sophisticated analysis of task descriptions, potentially leading to more accurate predictions of task automatability. We also utilize three task description datasets: O*NET Task statements \cite{onet}, ESCO Skills \cite{esco}, and Australian Labour Market Insights Tasks \cite{australian_labor_outlook}. We annotate each task with one of three labels: ``Substitution," ``Complementarity," or ``Negligibility," representing its automation susceptibility based on expert assessments. Using these annotations as ground truth, we apply a BERT-based classifier model \cite{devlin2018bert} to predict task automatability, benefiting from BERT's context understanding and semantic capabilities.\par

We extend our analysis to occupation and industry levels, using aggregated task-level predictions to assess overall automation susceptibility. This comprehensive view can reveal trends not observable at the task level, providing additional automation impact insights. Our research holds significant implications for workers, businesses, and policymakers by offering clarity on automation-susceptible tasks and demonstrating BERT's utility in labor economics.\par
The contributions of our research to the domain of automation impact on the labor market can be summarized as follows:\par
\begin{itemize}
\item Presents a novel approach to predict the automatability at task level which provides a more detailed and nuanced understanding of the impact of automation.
\item This work substantiates the hypothesis that task automatability can be discerned through mining the contextual and semantic information of textual task statements by leveraging BERT's contextual and semantic understanding capabilities.
\item Fine-tuned the BERT-based classifier on task statement data, outperforms traditional machine learning and neural network models, offering a robust tool for assessing automation's future impact on the labor market.
\item This research combine and annotates three public datasets: O*NET Task Statements, ESCO Skills, and Australian Labour Market Insights Tasks, creating a valuable resource that underscores the robustness of the BERT-based classifier across diverse data types.
\end{itemize}

\section{Related Work}
The body of literature on the automatability of occupations is vast and growing, with numerous studies focusing on predicting the susceptibility of entire occupations to automation. For instance, Frey and Osborne \cite{freyosborne} pioneered the field with their seminal work, using a Gaussian process classifier to estimate the probability of computerisation for 702 detailed occupations and concluded that about 47\% occupations in US is at high risk. Arntz \cite{arntz} further expanded on this by considering the heterogeneity of jobs within occupations and obtained a less alarming rate of 9\%, albeit their focus remained on the occupation level. These studies, while groundbreaking, have been critiqued for their broad-brush approach, which overlooks the granularity of tasks within occupations.\par

In parallel, there has been a surge of interest in the application of BERT-based models for sentence classification tasks. Devlin \cite{devlin2018bert} introduced BERT, a transformer-based model that leverages context in both directions, making it highly effective for understanding the semantics of text. Since then, BERT has been widely used in various natural language processing tasks, including sentence classification \cite{croce2020ganbert, lu2020vgcnbert}, and cross domain applications \cite{lavi2021consultantbert, xue2019finetuning} demonstrating its ability to capture contextual and semantic information effectively. However, this is limited research in automatability prediction leveraging sentences' contextual and semantic information and multiple studies have formulated hypothesis that the automatability of tasks is fundamentally a question of understanding the nature and requirements of the task which is often embedded in the contextual and semantic information of textual task statements \cite{towards, webb_2019_the}.\par

Our work bridges these two streams of literature. While we build upon the existing body of work on occupation automatability, we shift the focus from the occupation level to the task level. We argue that this provides a more nuanced and accurate picture of the potential impact of automation on the workforce. Furthermore, we leverage the power of BERT for sentence classification, applying it to the analysis of task descriptions to predict their automatability. This represents a novel application of BERT, extending its use beyond traditional natural language processing tasks. In doing so, we hope to contribute a new perspective to the ongoing discourse on the future of work in the age of automation.\par
\section{Method}
\begin{figure}[htp!]
\centerline{\includegraphics[width=1\columnwidth]{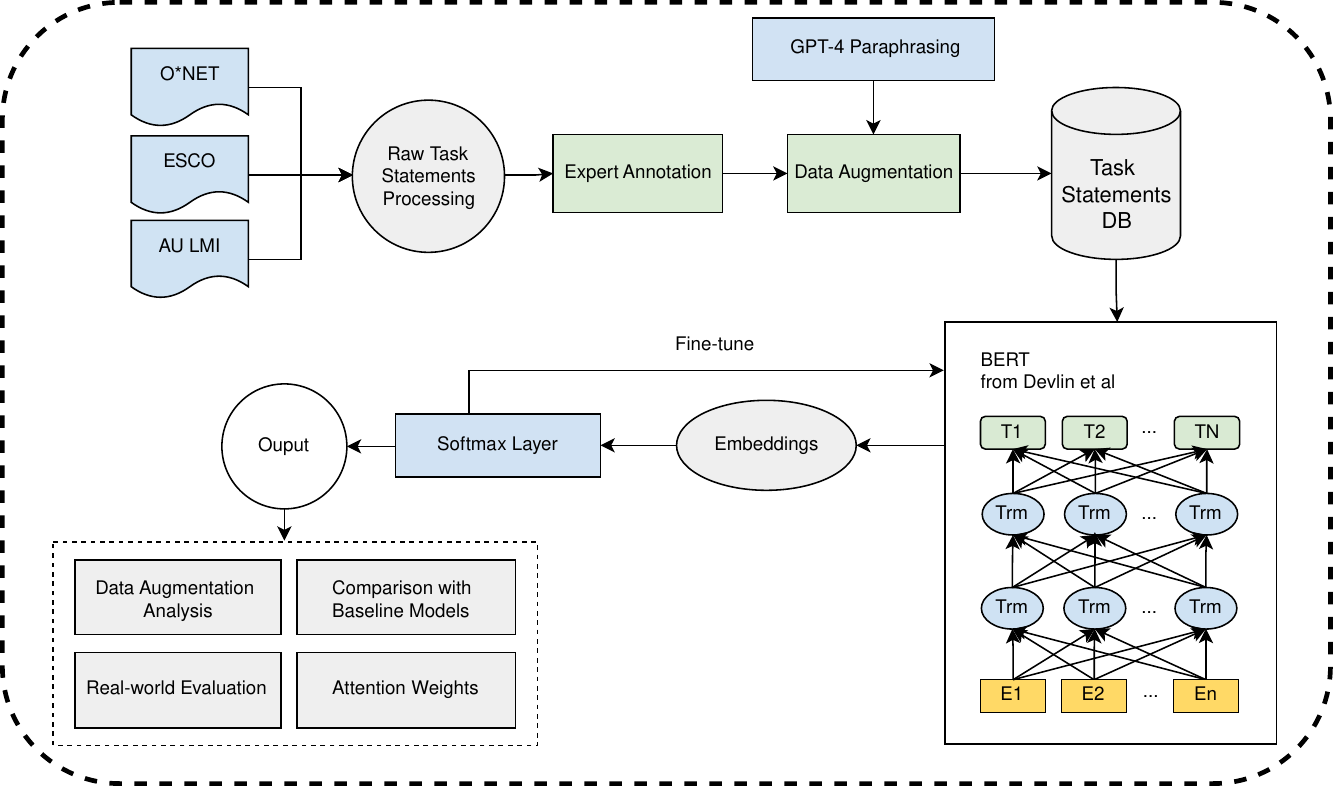}}
\caption{The framework of the proposed method.}
\label{fig: framework}
\end{figure}

In this section, we outline the whole framework (Fig. \ref{fig: framework}) employed to predict task level automatability leveraging a BERT-based classifier. Our approach first harnesses the BERT architecture to generate task statement embeddings, then construct a BERT-based classifier with a softmax layer for categorizing automatabilities. The model undergoes fine-tuning, tailored to the downstream classification task, optimizing its performance on the specific domain.

\subsection{Dataset Creation}
We used three public datasets: ONET Task Statements, ESCO skills, and AU Labor Market Insights Task statements. They provide a wide range of occupation-specific task statements with varying geographical and vocational coverage. We randomly sampled tasks: 5,060 from ONET, 4,783 from ESCO, and 3,356 from AU Labor Market Insights. This was done to ensure diversity across occupations, sectors, and regions for generalizable findings and to optimize computational resources and model training efficiency due to the high-dimensionality of text data and the large dataset sizes.
\subsubsection{Expert Annotation}
Our annotation process employs a voting mechanism involving five experts. For each task and each class, if a class receives more than three votes, it is designated as the final label for that task. This approach ensures a consensus among experts and enhances the reliability of the labels. Our expert annotation process labeled each task as ``Substitution" (fully automatable), ``Complementarity" (partial automation with human involvement), or ``Negligibility" (unlikely to be automated). This provides a categorical understanding of task-level automation susceptibility. We base our process on the Skill-Biased Technological Change (SBTC) \cite{card2002skill} and Routine-Biased Technological Change (RBTC) \cite{autor2003rbtc} hypotheses, which explain technological change's impact on labor markets. SBTC postulates that technology favors high-skill tasks and replaces lower-skilled routine jobs \cite{kristal2020computerization}, while RBTC asserts that technology replaces routine tasks, regardless of skill level \cite{buyst2018job}. We also incorporate six major automation bottlenecks : Complex Problem Solving, Social Interaction and Emotional Intelligence, Fine Motor Skills and Dexterity, Creative and Artistic Abilities, Contextual Understanding, and Ethical Decision Making \cite{chui2016where, nedelkoska2018automation}. These bottlenecks highlight areas where humans still outperform machines.

\subsubsection{Data Augmentation}
The initial exploration of the labeled data revealed a class imbalance issue, a common problem in many real-world classification tasks \cite{krawczyk2016learning}. This imbalance can lead to biased models, overfitting to the majority class and neglecting minority ones. Conventional data augmentation methods suitable for image data aren't applicable to text data due to its sequential and semantic nature \cite{zhang2015character}. Apart from class imbalance issue, effectively increasing the volume of training data is particularly beneficial for deep learning models like BERT that perform better with larger datasets.\par
We addressed this using GPT-4 \cite{openai2023gpt4} to paraphrase sentences, creating new task statements with identical meanings but different wording. GPT-4's ability to generate high-quality and varied paraphrases ensures the augmented data remains relevant and increases the classifier's robustness to different task expressions. Specifically, each statement was fed into GPT-4 for paraphrasing, followed by automated checks and expert reviews to validate the paraphrases for semantic similarity to the original statements:\par 
\noindent\textbf{Original Sentence:}\par
\noindent Generate reports utilizing visual aids such as charts, graphs, and narratives by examining and documenting test data.\par
\noindent\textbf{Paraphrased Sentence:}\par
\noindent \textit{Create} reports that \textit{incorporate} visual \textit{elements} like \textit{diagrams}, \textit{plots}, and descriptive narratives by \textit{scrutinizing} and \textit{recording} the results of tests.

\subsection{BERT-based Classifier}
BERT stands out with its multi-layer bidirectional Transformer encoder, which enables it to capture the deep contextual information embedded in text data \cite{devlin2018bert}. Mathematically, this is achieved by applying the attention mechanism, where the output embeddings \textit{E} for a given input token is a weighted sum of all input token embeddings:
\begin{equation}
E = \sum \text{{\textit{Attention\_Scores}}} \times \text{{\textit{Token\_Embeddings}}}
\end{equation}
For our classification task, we fine-tuned a pre-trained BERT model on our task statement data. Each task statement was fed into BERT, which transformed the text into high-dimensional embeddings.\par

The embeddings were then passed through a softmax function to generate the final class probabilities. If we denote the output of our model before the softmax layer as \(z\), the softmax function can be expressed as:
\begin{equation}
\text{{\textit{Softmax}}}(z_i) = \frac{e^{z_i}}{\sum e^{z_j}}
\end{equation}
where \(z_i\) is the output corresponding to the i-th class and the sum in the denominator runs over all possible classes.\par
The training process involved adjusting the model's parameters to minimize the discrepancy between the predicted and actual class labels, typically quantified using the cross-entropy loss function:
\begin{equation}
L = -\sum \text{{\textit{y\_true}}} \times \log(\text{{\textit{y\_pred}}})
\end{equation}
where \textit{y\_true} is the true label and \textit{y\_pred} is the predicted probability.
\subsection{Attention Mechanism}
A fundamental feature of BERT's ability to capture deep contextual information is its use of the attention mechanism, specifically, the scaled dot-product attention as introduced in the original Transformer model \cite{vaswani2017attention}. This attention mechanism allows the model to weigh the importance of different words in a sentence, providing a powerful tool for understanding language semantics.\par
Mathematically, the attention mechanism can be described as mapping a query and a set of key-value pairs to an output. Given a query (\textit{Q}), keys (\textit{K}), and values (\textit{V}), the output (\textit{O}) of the attention mechanism is calculated as a weighted sum of the values, where the weight assigned to each value is determined by the query's compatibility with the corresponding key:
\begin{equation}
\text{\textit{Attention}}(Q, K, V) = \text{\textit{softmax}}\left(\frac{QK^T}{\sqrt{d_k}}\right)V
\end{equation}
where the denominator \(\sqrt{d_k}\) is used for scaling, with \(d_k\) being the dimension of the key vectors. The softmax function ensures the attention scores are normalized to lie between 0 and 1, thus can be interpreted as probabilities. This results in words that are more important to the meaning of the sentence receiving a higher attention score, while less important words receive a lower score.

The use of attention weights allows us to visualize and interpret the model's decision-making process. By examining these weights, we can understand which parts of the input sentence are most influential in determining the output of the model.

\section{Experiment}
\subsection{Data}
The datasets were partitioned into training, evaluation, and test subsets following an 8:1:1 ratio, providing a comprehensive and balanced basis for our experimental studies. And both the original and augmented datasets would be passed through the model to evaluate and quantify the influence of data augmentation.\par
The table \ref{overviewdata} presented below provides a visual representation of both the initial and augmented datasets. Specifically, during the training phase using the original dataset, we observed that the AU LMI dataset exhibited superior performance compared to the other two datasets. Consequently, during the process of data augmentation, we prioritized augmenting the ONET and ESCO datasets to a greater extent, while applying relatively fewer augmentations to the AU LMI dataset. This decision was based on the recognition that ONET and ESCO present greater challenges in terms of machine learning, warranting additional exposure to augmented instances for effective learning.
\begin{table}[htbp]
\caption{The overview of dataset}
\label{overviewdata}
\centering
\begin{tblr}{
  cell{1}{1} = {c=2}{},
  cell{2}{1} = {r=2}{},
  cell{4}{1} = {r=2}{},
  cell{6}{1} = {r=2}{},
  vline{1,2,3,4,5,6},
  hline{1,2,3,4,5,6,7,8}
}
                &           & O*NET & ESCO & AU LMI \\ 
Substitution    & Original  & 1,594  & 1,435 & 998    \\ 
                & \textbf{Augmented} & \textbf{3,188}  & \textbf{3,157} & \textbf{1,796}   \\
Complementarity & Original  & 2,519  & 2,272 & 1,776   \\ 
                & \textbf{Augmented} & \textbf{3,023}  & \textbf{3,181} & \textbf{1,776}   \\
Negligibility  & Original  & 947   & 1,076 & 582    \\ 
                & \textbf{Augmented} & \textbf{3,030}  & \textbf{3,228} & \textbf{1,764} 
\end{tblr}
\end{table}

\subsection{Evaluation Metrics}
We assessed our model and baselines using precision, recall, and F1 score. Precision evaluates the ratio of correct positive predictions to all positive predictions, indicating the model's false-positive avoidance. Recall measures the fraction of true positives from all actual positives, reflecting the model's ability to recognize all relevant instances. The F1 score, the harmonic mean of precision and recall, balances these aspects—a high F1 score indicates a model with high precision and recall, and vice versa.\par
In our multi-class problem, these metrics were computed for each class, considering the class in question as positive and the rest as negative. The average of per-class metrics provided a single performance measure across all classes.

\subsection{Baselines}
In our work, we make a comprehensive comparison of our proposed BERT-based model with a range of baseline models, encompassing both traditional machine learning approaches, neural network architectures, and other transformer models. These baselines span different approaches to text classification, allowing us to assess the relative merits of our approach in a broad context.

For traditional classifiers, we employ Logistic Regression \cite{cessie1992ridge}, Random Forest \cite{breiman2001random}, and Support Vector Machines (SVM) \cite{cortes1995support}. These models, with their varying theoretical underpinnings, provide a solid foundation against which to compare our more complex neural model. They represent different forms of linear and non-linear decision boundaries and incorporate different forms of regularization and ensemble learning.

In the realm of neural networks, we utilize Bi-directional Long Short-Term Memory (BiLSTM) \cite{graves2005framewise}, Gated Recurrent Units (GRU) \cite{cho2014learning}, and one-dimensional Convolutional Neural Networks (Conv1D) \cite{yoon2020convolutional}. These architectures demonstrate the potential of deep learning for text classification, and their comparison with our model allows us to quantify the value of pre-training and transformer architecture in this context.

Finally, we include three additional transformer models in our baselines: ALBERT \cite{lan2019albert}, ELECTRA \cite{clark2020electra}, and DistilBERT \cite{sanh2019distilbert}. As close relatives of BERT, they provide a stringent test of the specific benefits of our BERT-based approach. By contrasting with these models, we can highlight the unique strengths of our chosen methodology.

\subsection{Results}
The table \ref{results} summarizes the results for proposed BERT and other baseline models of different datasets. We evaluate them with precision, recall and F1-score.\par

\begin{table*}
\caption{The results of proposed model and baselines.}
\label{results}
\centering
\begin{tblr}{
  cell{1}{1} = {r=2}{c},
  cell{1}{2} = {c=3}{c},
  cell{1}{5} = {c=3}{c},
  cell{1}{8} = {c=3}{c},
  cell{1}{11} = {c=3}c{},
  cell{16}{12} = {c=2}{c},
  vlines,
  hlines,
  column{1} = {c},
  column{2-13} = {c}
}
\textbf{Model}   & \textbf{O*NET}     &        &        & \textbf{ESCO}      &        &        & \textbf{AU LMI}    &        &        & \textbf{O*NET + ESCO + AU LMI} &        &        \\
        & \textbf{Precision} & \textbf{Recall} & \textbf{F1}     & \textbf{Precision} & \textbf{Recall} & \textbf{F1}     & \textbf{Precision} & \textbf{Recall} & \textbf{F1}     & \textbf{Precision}             & \textbf{Recall} & \textbf{F1}     \\
LR      & 0.6091    & 0.61   & 0.609  & 0.5915    & 0.5915 & 0.5911 & 0.657     & 0.6602 & 0.6566 & 0.5902                & 0.5922 & 0.5908 \\
SVM     & 0.6087    & 0.6072 & 0.6046 & 0.5738    & 0.5719 & 0.5707 & 0.6462    & 0.6473 & 0.6452 & 0.5837                & 0.5867 & 0.5811 \\
RF      & 0.5922    & 0.593  & 0.5921 & 0.5657    & 0.5656 & 0.5652 & 0.6449    & 0.6483 & 0.6399 & 0.5889                & 0.5909 & 0.5891 \\
BiLSTM  & 0.7487    & 0.7562 & 0.7532 & \textbf{0.7973}    & \textbf{0.7898} & \textbf{0.7876} & \textbf{0.8997}    & \textbf{0.8994} & \textbf{0.8995} & 0.7676                & 0.7701 & 0.7665 \\
GRU     & \textbf{0.7712}    & \textbf{0.7676} &\textbf{ 0.7712} & 0.7587    & 0.7612 & 0.7611 & 0.8677    & 0.8701 & 0.8664 & 0.7609                & 0.7624 & 0.7611 \\
CONV1D  & 0.7695    & 0.7742 & 0.7778 & 0.7745    & 0.7812 & 0.7822 & 0.8424    & 0.8402 & 0.8468 & 0.7766                & 0.7798 & 0.7723 \\
ALBERT  & 0.7703    & 0.6989 & 0.6988 & 0.7687    & 0.7723 & 0.7701 & 0.7997    & 0.8033 & 0.7946 & 0.7253                & 0.7114 & 0.6801 \\
ELECTRA & 0.7544    & 0.7621 & 0.7602 & 0.7384    & 0.7381 & 0.7285 & 0.8145    & 0.8093 & 0.8122 & 0.7459                & 0.7419 & 0.7425 \\
\textbf{BERT}    & 0.7698    & 0.7643 & 0.7594 & 0.7701    & 0.7723 & 0.7698 & 0.8241    & 0.8225 & 0.8198 & \textbf{0.7981}                & \textbf{0.7955} & \textbf{0.7944}   
\end{tblr}
\end{table*}

As we can see from the results, the GRU model demonstrates the highest performance on the O*NET dataset, with Precision, Recall, and F1-Score at 0.7712, 0.7676, and 0.7712, respectively. BiLSTM exhibits superior performance on the ESCO dataset, with Precision, Recall, and F1-Score at 0.7973, 0.7898, and 0.7876, respectively. The same model outperforms others on the AU LMI dataset, with all three metrics around 0.8995.\par

When all datasets are combined, our proposed BERT-based classifier emerges as the most effective model. The precision, recall, and F1-score are 0.7981, 0.7955, and 0.7944, respectively, which are the highest among all models. These results indicate the robustness of BERT in handling diverse and large-scale datasets and its superiority over traditional, neural network, and other transformer models for this multi-class classification task. We will use BERT as the best performance model for follow-up experiments and analysis.

\subsection{Data Augmentation Analysis}
\begin{figure}[htp!]
\centerline{\includegraphics[width=0.8\columnwidth]{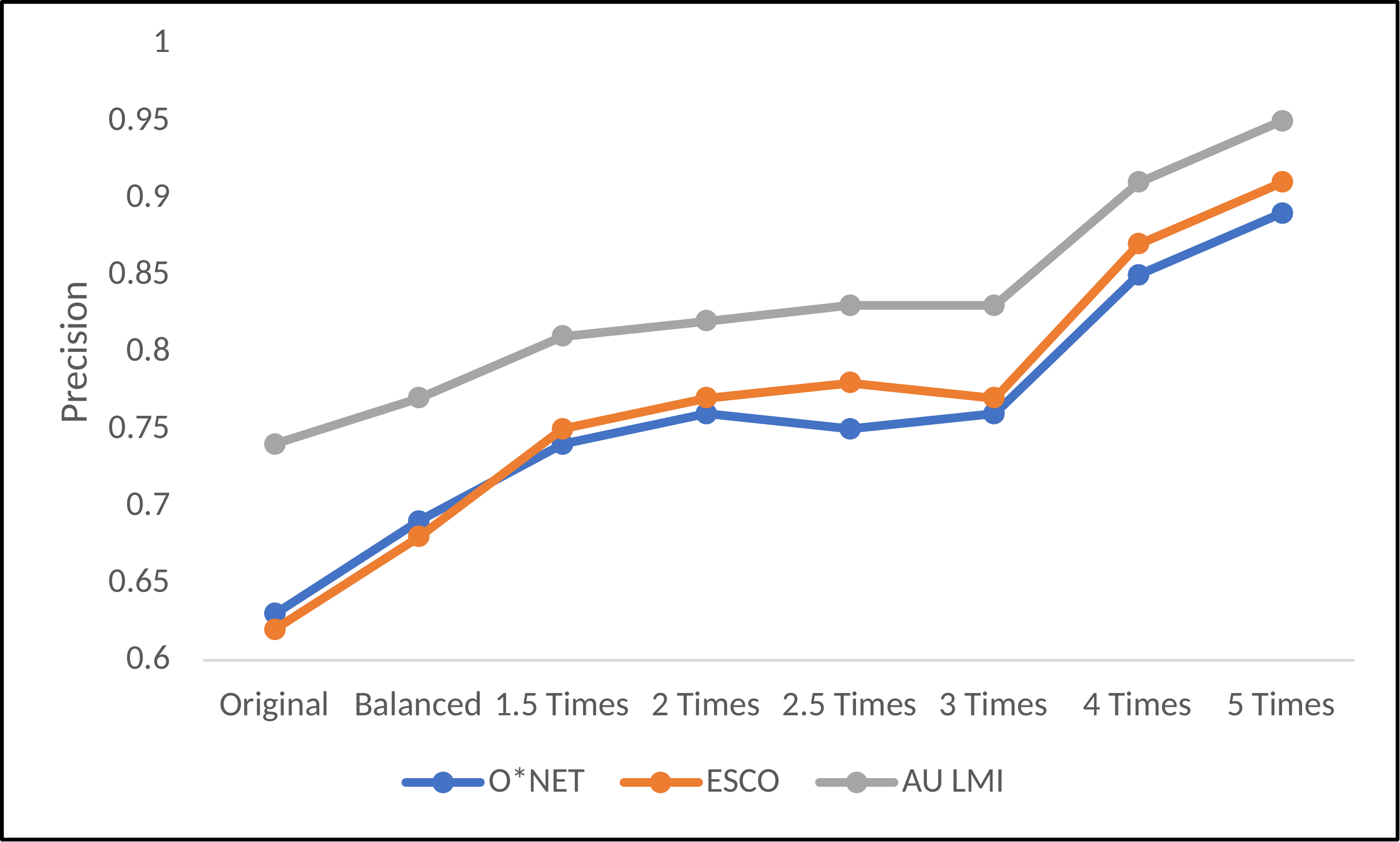}}
\caption{The Comparison of different augmentation.}
\label{fig: augmentation}
\end{figure}

In this section, we examine the impact of different augmentation levels on our BERT model's performance. The data augmentation methods include no augmentation (Original), augmenting data to match the class with the largest instances (Balanced), and incremental augmentation of the original data by 1.5, 2, 2.5, 3, 4, and 5 times. The results are shown below as Fig. \ref{fig: augmentation}:\par
A careful review of the results indicates that augmentation improves the precision of the model across all three datasets (ONET, ESCO, and AU LMI). For the ONET and ESCO datasets, the precision increases modestly from ``Balanced" to ``2 Times" and then plateaus slightly at ``2.5 Times". However, a dramatic rise is seen at ``4 Times", implying potential over-augmentation, which could lead to an overfitted model.\par
Therefore, we choose to augment our data between ``Balanced" and ``2 Times" for the O*NET and ESCO datasets. As for the AU LMI dataset, which already shows good performance, we only balance the classes.

\subsection{Model Robustness on Different Data Split}
\begin{figure}[htp!]
\centerline{\includegraphics[width=0.8\columnwidth]{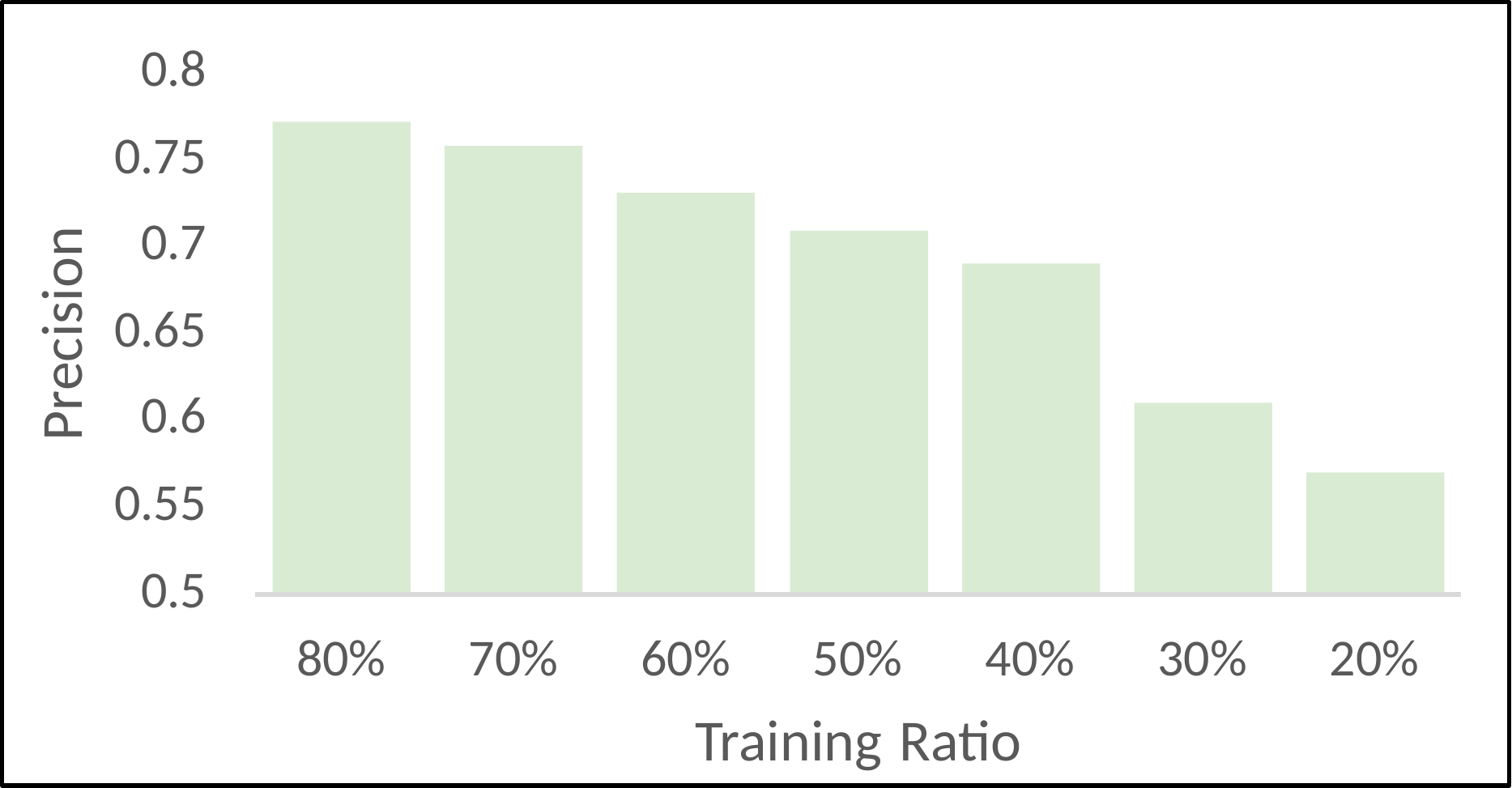}}
\caption{The performance of model at different data split.}
\label{fig: robustness}
\end{figure}

To ascertain the robustness of our proposed model for real-world applications, we conducted an experiment with varying training set sizes. The dataset was shuffled, and training sets were created with different proportions of the original dataset, namely 80\%, 70\%, 60\%, 50\%, 40\%, 30\%, and 20\%. The remaining data in each scenario was used for performance evaluation. The results are presented at Fig. \ref{fig: robustness}:\par
As can be observed, the proposed BERT model's performance is relatively stable, and it gets better performance with the increase of training data ratio. Besides, our model with only 20\% training data can catch up with the baselines who use 80\% data for training, which validates the effectiveness of BERT in the task of predicting the automatability at task level.

\subsection{Attention Weights Visualization}
For the visualization of the model's attention, we have employed the technique of Wordcloud.
For class Substitution, the prominent terms include ``using system", ``machinery operate", ``data record", ``routine perform", and ``trucks load". These tasks are generally routine and predictable, and hence, are more prone to automation. On the other hand, for class Complementarity and class Negligibility, the key terms hint at tasks that require a higher level of human judgment or interaction like ``information provide", ``educational program", ``research conduct", ``medical procedures", ``human expertise", and ``children care". These tasks correlate with recognized automation bottlenecks, corroborating the model's predictions. The results can be found at Fig. \ref{fig: wordcloud}:
\begin{figure}[htp!]
\centerline{\includegraphics[width=0.8\columnwidth]{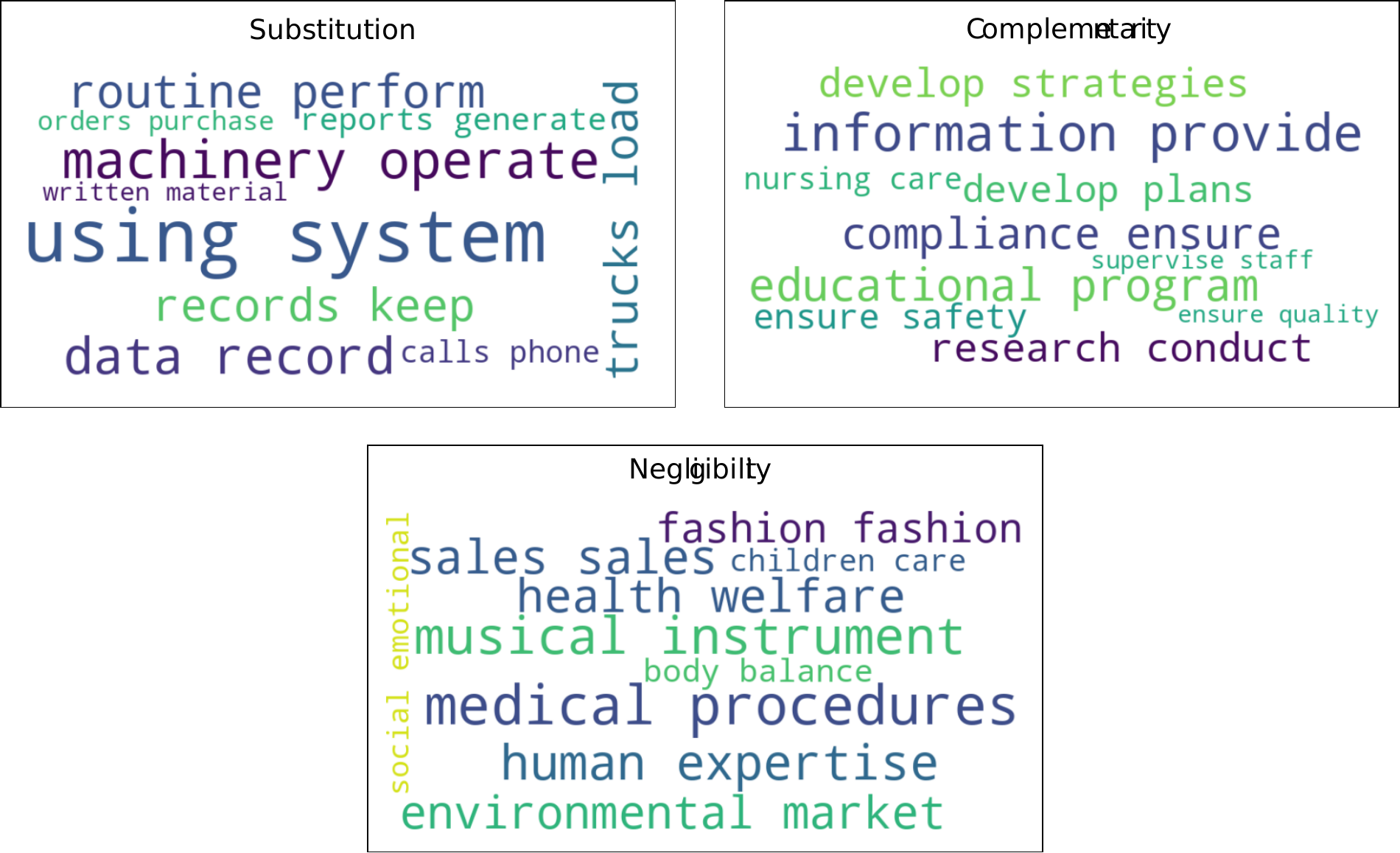}}
\caption{The Wordclouds for different categories.}
\label{fig: wordcloud}
\end{figure}

\section{Results and Discussion}
\begin{figure}[htp!]
\centerline{\includegraphics[width=0.8\columnwidth]{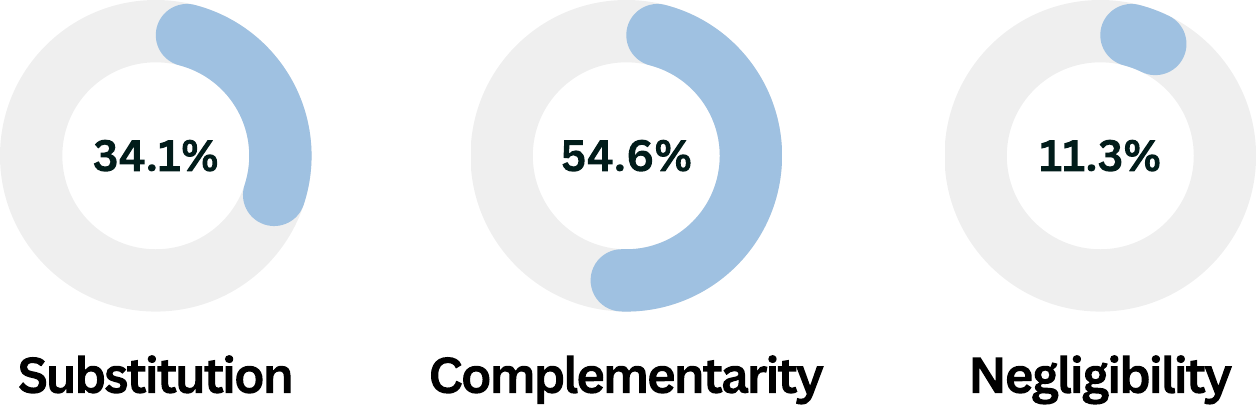}}
\caption{The Distribution of O*NET Task Automatability.}
\label{fig: distrubution}
\end{figure}

Having established the optimal performance of our BERT model through a series of experiments, we moved to a practical application, leveraging the model for inference on real-world datasets. We selected O*NET task statements, a rich and diverse dataset that encapsulates a broad variety of professional tasks, totalling 19,530 individual tasks. The results showed that among the total tasks, 6664 tasks were labeled as ``Substitution", 10,678 tasks were ``Complementarity" and 2188 tasks were ``Negligibility" which could be visualized as Fig. \ref{fig: distrubution}:\par
\subsection{Assessment of Automatability at Occupation Level}
\begin{figure}[htp!]
\centerline{\includegraphics[width=0.8\columnwidth]{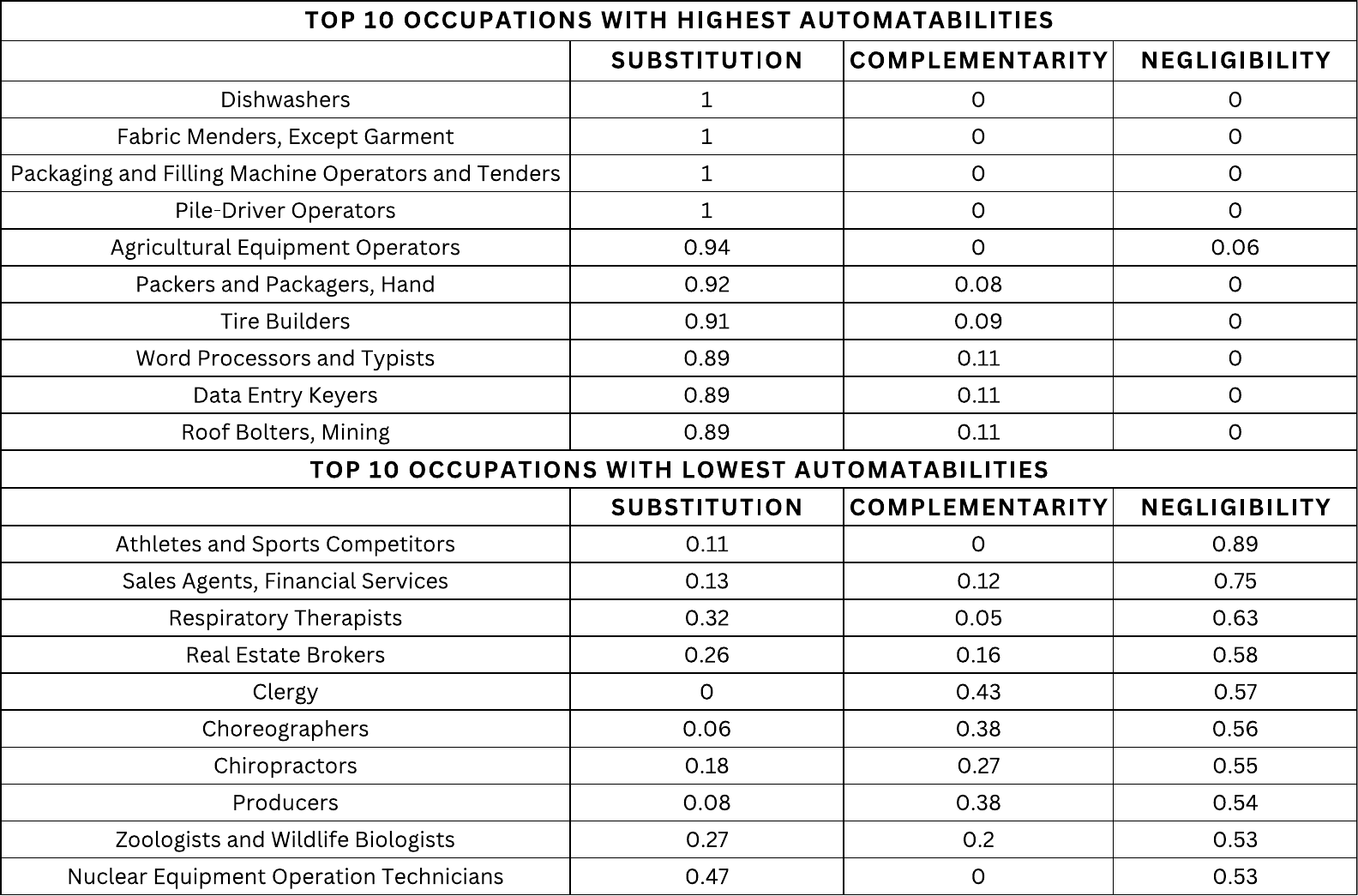}}
\caption{Top 10 occupations that have highest and lowest automatabilities.}
\label{fig: top10}
\end{figure}

We mapped individual task automatability measures to 974 distinct occupations using the O*NET task statement to occupation mapping. This allowed us to assess automatability at an occupation level by aggregating tasks and quantifying task type distributions. The results, summarized in Table \ref{fig: top10}, show the top 10 occupations with the highest substitution and negligibility, indicating the most and least likely automated occupations, respectively.
As can be seen from the results, occupations with high automation susceptibility are primarily manual labor, repetitive tasks, or data-intensive roles, such as "Dishwashers" and "Packaging and Filling Machine Operators and Tenders". Conversely, roles requiring human interaction, creativity, specialized skills, or unpredictable environments, like "Athletes and Sports Competitors" and "Respiratory Therapists", showed lower automatability. \par
Our research findings indicate that out of 974 ONET occupations, 244 display a substitution score exceeding 50\%. This suggests that approximately 25.1\% of occupations within the ONET database are at substantial risk of automation. This outcome is aligned with the findings put forth in \cite{dawei}, thus adding robustness to our estimations. Furthermore, these results are nested between the estimations reported in \cite{freyosborne} which places 47\% of occupations at risk and \cite{arntz} which estimates the figure at a comparatively lower 9\%. Considering the limitations often associated with automated risk evaluations at the occupation and job level, the median value that our research arrives at seems reasonable. This is due to our methodology which leverages granular task statement data to predict occupation-level risks.\par
Additionally, our results indicate that a majority of occupations, specifically 603 out of 974 which is 61.8\%, face a high risk of complementarities. Conversely, a relatively smaller fraction, constituting 128 occupations or 13.1\%, appear to be comparatively safe from impending automation over the forthcoming decades. These conclusions underscore the nuanced complexities underpinning automation risks and the need for more granular data analysis in this domain.
\subsection{Assessment of Automation Vulnerability Among Industries}
\begin{figure}[htp!]
\centerline{\includegraphics[width=1\columnwidth]{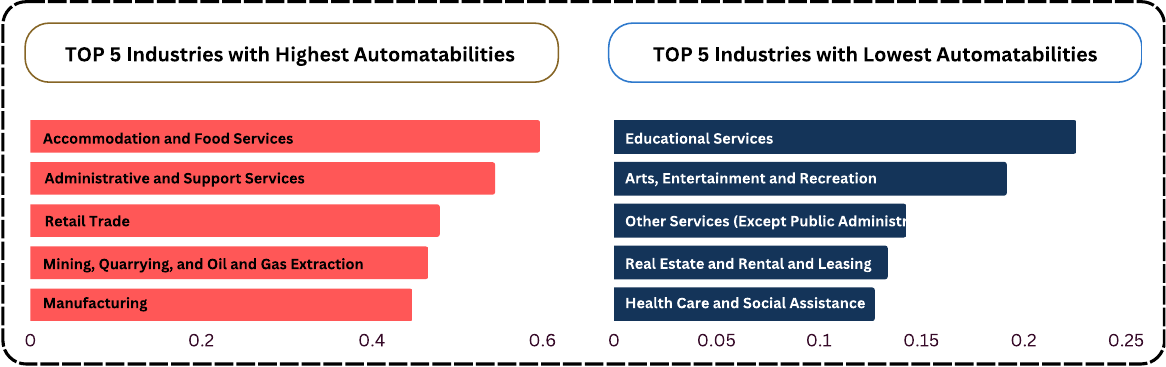}}
\caption{Industries with highest and lowest automatabilities.}
\label{fig: top5}
\end{figure}

Building on our earlier findings of automatability at the occupation level, we bridge the occupation-industry gap utilizing O*NET's detailed occupation-industry mapping. Our results shown as Fig. \ref{fig: top5} indicate a diverse spectrum of automation vulnerability across sectors. On one end, industries such as ``Accommodation and Food Services", ``Administrative and Support Services", ``Retail Trade", ``Mining, Quarrying, and Oil and Gas Extraction", and ``Manufacturing" emerge as the sectors most susceptible to automation. These industries comprise occupations with high automatability scores, revealing their high likelihood of experiencing significant changes due to automation.\par
On the other hand, industries such as ``Educational Services", ``Arts, Entertainment and Recreation", ``Other Services (Except Public Administration)", ``Real Estate and Rental and Leasing", and ``Health Care and Social Assistance" stand on the less vulnerable end of the automation spectrum. Occupations within these sectors possess low automatability scores, indicating a lower likelihood of their roles being fully automated, largely due to the complexity of tasks or the high level of human judgment, interaction, and creativity required.\par

To validate the results, we leveraged the insights from the \cite{mckinsey2017jobs}, and identified 4 sectors showing siginifican susceptibility to automation which are ``Accommodation and Food Services", ``Manufacturing", ``Retail Trade", and ``Transportation and Warehousing". These findings align well with our own results.

\section{Conclusion}
This study presents a unique approach to predict task-level automatability with a BERT-based classifier, utilizing three diverse, public datasets and expert annotations. Rigorous experiments demonstrated our model's efficacy.\par
In practical application, we applied our model to real-world datasets, providing a comprehensive perspective of occupational automatability. Our findings indicate that approximately 25.1\% of occupations within the O*NET database are at substantial risk of automation. Furthermore, our results reveal a diverse spectrum of automation vulnerability across sectors, with industries such as ``Accommodation and Food Services", ``Administrative and Support Services", ``Retail Trade", ``Mining, Quarrying, and Oil and Gas Extraction", and ``Manufacturing" emerging as the sectors most susceptible to automation.\par
These findings have significant implications for workers and policymakers. For workers, understanding the susceptibility of their tasks to automation can help them make informed decisions about their career paths and upskilling opportunities. For policymakers, these insights can guide the development of policies and initiatives aimed at managing the transition to an increasingly automated workforce. This could include strategies for retraining and reskilling workers, as well as measures to support industries and regions that are particularly vulnerable to automation.\par
In conclusion, our research provides a robust and effective approach to predicting task-level automatability, offering valuable insights for policymakers, educators, and workers. By understanding the potential impact of automation on different tasks, occupations, and industries, we can better prepare for the future of work.\par

\section*{Acknowledgment}
This work is supported by the Australian Research Council (ARC) under Grant No. DP220103717, and LE220100078.





\onecolumn
\begin{appendices}
\clearpage


\section*{Supplementary material}

APPENDIX A
THE RESULTS OF OCCUPATIONS' AUTOMATABILITY

| Occupation | Substitution | Complementarity | Negligibility |
|---|---|---|---|
| Attendants and Bartender Helpers | 1 | 0 | 0 |
| Dishwashers | 1 | 0 | 0 |
| Fabric Menders, Except Garment | 1 | 0 | 0 |
| Operators and Tenders | 1 | 0 | 0 |
| Pile-Driver Operators | 1 | 0 | 0 |
| Agricultural Equipment Operators | 0.941176471 | 0 | 0.058823529 |
| Packers and Packagers, Hand | 0.916666667 | 0.083333333 | 0 |
| Tire Builders | 0.909090909 | 0.090909091 | 0 |
| Material Movers, Hand | 0.904761905 | 0.095238095 | 0 |
| Fence Erectors | 0.9 | 0.05 | 0.05 |
| Word Processors and Typists | 0.894736842 | 0.105263158 | 0 |
| Data Entry Keyers | 0.888888889 | 0.111111111 | 0 |
| Roof Bolters, Mining | 0.888888889 | 0.111111111 | 0 |
| Prepress Technicians and Workers | 0.875 | 0 | 0.125 |
| Maids and Housekeeping Cleaners | 0.869565217 | 0.130434783 | 0 |
| Underground Mining | 0.866666667 | 0.066666667 | 0.066666667 |
| Operators, Except Postal Service | 0.862068966 | 0.137931034 | 0 |
| Statistical Assistants | 0.857142857 | 0.071428571 | 0.071428571 |
| Retail Sales | 0.857142857 | 0.142857143 | 0 |
| Hand | 0.85 | 0.15 | 0 |
| Medical Assistants | 0.85 | 0.15 | 0 |
| Foundry Mold and Coremakers | 0.846153846 | 0.153846154 | 0 |
| Tire Repairers and Changers | 0.846153846 | 0.153846154 | 0 |
| Parking Lot Attendants | 0.833333333 | 0.111111111 | 0.055555556 |
| Auditing Clerks | 0.833333333 | 0.166666667 | 0 |
| Credit Authorizers | 0.833333333 | 0.166666667 | 0 |
| and Tenders | 0.823529412 | 0.117647059 | 0.058823529 |
| Helpers--Production Workers | 0.823529412 | 0.176470588 | 0 |
| Pharmacy Aides | 0.823529412 | 0.176470588 | 0 |
| Cooks, Short Order | 0.818181818 | 0.181818182 | 0 |
| Operators | 0.818181818 | 0.181818182 | 0 |
| Applicators, Vegetation | 0.8 | 0.2 | 0 |
| Machine Setters, Operators, and Tenders | 0.791666667 | 0.166666667 | 0.041666667 |
| Cooks, Fast Food | 0.789473684 | 0.105263158 | 0.105263158 |
| Baristas | 0.789473684 | 0.210526316 | 0 |
| File Clerks | 0.789473684 | 0.210526316 | 0 |
| Food Servers, Nonrestaurant | 0.785714286 | 0.071428571 | 0.142857143 |
| Compacting Machine Setters, Operators, and Tenders | 0.785714286 | 0.107142857 | 0.107142857 |
| Locksmiths and Safe Repairers | 0.785714286 | 0.214285714 | 0 |
| Bakers | 0.777777778 | 0.222222222 | 0 |

| Occupation | | | |
|---|---|---|---|
| Marking Clerks | 0.777777778 | 0.222222222 | 0 |
| Technicians | 0.777777778 | 0.222222222 | 0 |
| Operators, and Tenders, Metal and Plastic | 0.769230769 | 0.230769231 | 0 |
| Machine Feeders and Offbearers | 0.769230769 | 0.230769231 | 0 |
| Signal and Track Switch Repairers | 0.769230769 | 0.230769231 | 0 |
| Statement Clerks | 0.769230769 | 0.230769231 | 0 |
| Equipment Assemblers | 0.764705882 | 0.235294118 | 0 |
| Precipitating, and Still Machine Setters, Operators, and Tenders | 0.761904762 | 0.19047619 | 0.047619048 |
| Service Attendants | 0.75 | 0.25 | 0 |
| Medical Secretaries | 0.75 | 0.25 | 0 |
| Pharmacy Technicians | 0.75 | 0.25 | 0 |
| Collectors | 0.75 | 0.25 | 0 |
| Solderers and Brazers | 0.75 | 0.25 | 0 |
| Stock Clerks, Sales Floor | 0.75 | 0.25 | 0 |
| Cashiers | 0.740740741 | 0.222222222 | 0.037037037 |
| Processors, and Processing Machine Operators | 0.736842105 | 0.157894737 | 0.105263158 |
| Highway Maintenance Workers | 0.736842105 | 0.263157895 | 0 |
| Payroll and Timekeeping Clerks | 0.736842105 | 0.263157895 | 0 |
| Answering Service | 0.736842105 | 0.263157895 | 0 |
| Technicians | 0.733333333 | 0.2 | 0.066666667 |
| Food Preparation Workers | 0.733333333 | 0.233333333 | 0.033333333 |
| Tree Trimmers and Pruners | 0.730769231 | 0.230769231 | 0.038461538 |
| Finishers | 0.727272727 | 0.181818182 | 0.090909091 |
| Pickling Equipment Operators and Tenders | 0.727272727 | 0.272727273 | 0 |
| and Samplers, Recordkeeping | 0.727272727 | 0.272727273 | 0 |
| Computer | 0.722222222 | 0.166666667 | 0.111111111 |
| Machine Setters, Operators, and Tenders, Metal and Plastic | 0.71875 | 0.28125 | 0 |
| Maids and Housekeeping Cleaners | 0.714285714 | 0.285714286 | 0 |
| Searchers | 0.705882353 | 0.235294118 | 0.058823529 |
| Equipment Operators | 0.7 | 0.25 | 0.05 |
| Baking, and Drying Machine Operators and Tenders | 0.7 | 0.3 | 0 |
| Postal Service Clerks | 0.7 | 0.3 | 0 |
| Installers and Repairers | 0.7 | 0.3 | 0 |
| Machine Setters, Operators, and Tenders, Metal and Plastic | 0.696969697 | 0.303030303 | 0 |
| Booth Cashiers | 0.692307692 | 0.307692308 | 0 |
| Mine Shuttle Car Operators | 0.692307692 | 0.307692308 | 0 |

| Occupation | | | |
|---|---|---|---|
| Roustabouts, Oil and Gas | 0.692307692 | 0.307692308 | 0 |
| Tapers | 0.6875 | 0.125 | 0.1875 |
| Equipment Installers and | 0.684210526 | 0.263157895 | 0.052631579 |
| Order Clerks | 0.684210526 | 0.315789474 | 0 |
| Endoscopy Technicians | 0.666666667 | 0.083333333 | 0.25 |
| Rock Splitters, Quarry | 0.666666667 | 0.111111111 | 0.222222222 |
| Equipment | 0.666666667 | 0.19047619 | 0.142857143 |
| Correspondence Clerks | 0.666666667 | 0.238095238 | 0.095238095 |
| Operators, Refinery Operators, and Gaugers | 0.666666667 | 0.25 | 0.083333333 |
| Cooks, Institution and Cafeteria | 0.666666667 | 0.266666667 | 0.066666667 |
| Semiconductor Processors | 0.666666667 | 0.291666667 | 0.041666667 |
| Sheet Metal Workers | 0.666666667 | 0.291666667 | 0.041666667 |
| Tool Operators, Metal and Plastic | 0.666666667 | 0.296296296 | 0.037037037 |
| Glaziers | 0.666666667 | 0.296296296 | 0.037037037 |
| Computer Operators | 0.666666667 | 0.333333333 | 0 |
| Cutters and Trimmers, Hand | 0.666666667 | 0.333333333 | 0 |
| Electro-Mechanical Technicians | 0.666666667 | 0.333333333 | 0 |
| and Tenders | 0.666666667 | 0.333333333 | 0 |
| Workers | 0.666666667 | 0.333333333 | 0 |
| Operators, and Tenders, Wood | 0.666666667 | 0.333333333 | 0 |
| Library Assistants, Clerical | 0.65625 | 0.28125 | 0.0625 |
| Freight and Cargo Inspectors | 0.65 | 0.25 | 0.1 |
| Fitters | 0.65 | 0.3 | 0.05 |
| Gaming Cage Workers | 0.647058824 | 0.176470588 | 0.176470588 |
| Food Concession, and Coffee Shop | 0.647058824 | 0.352941176 | 0 |
| Brickmasons and Blockmasons | 0.642857143 | 0.285714286 | 0.071428571 |
| Tellers | 0.642857143 | 0.285714286 | 0.071428571 |
| Waiters and Waitresses | 0.64 | 0.36 | 0 |
| Plasterers, and Stucco Masons | 0.636363636 | 0.090909091 | 0.272727273 |
| Hunters and Trappers | 0.636363636 | 0.181818182 | 0.181818182 |
| Crane and Tower Operators | 0.636363636 | 0.272727273 | 0.090909091 |
| Clerks | 0.636363636 | 0.272727273 | 0.090909091 |
| Cooks, Private Household | 0.636363636 | 0.363636364 | 0 |
| Insurance Claims Clerks | 0.636363636 | 0.363636364 | 0 |
| Operators, and Tenders, Metal and Plastic | 0.633333333 | 0.366666667 | 0 |
| Operators and Tenders | 0.631578947 | 0.315789474 | 0.052631579 |
| Setters, Operators, and Tenders | 0.631578947 | 0.368421053 | 0 |
| Parts Salespersons | 0.631578947 | 0.368421053 | 0 |
| Medical Equipment Preparers | 0.625 | 0.25 | 0.125 |
| Couriers and Messengers | 0.625 | 0.3125 | 0.0625 |
| Technologists | 0.625 | 0.3125 | 0.0625 |

| Occupation | | | |
|---|---|---|---|
| Dental Assistants | 0.625 | 0.375 | 0 |
| Assistants, Except Legal, Medical, and Executive | 0.625 | 0.375 | 0 |
| Telephone Operators | 0.625 | 0.375 | 0 |
| Treatment Plant and System | 0.625 | 0.375 | 0 |
| Machine Setters, Operators, and Tenders | 0.620689655 | 0.137931034 | 0.24137931 |
| Setters, Operators, and Tenders, Synthetic and Glass Fibers | 0.619047619 | 0.333333333 | 0.047619048 |
| Ticket Agents and Travel Clerks | 0.619047619 | 0.333333333 | 0.047619048 |
| Machine Operators | 0.619047619 | 0.380952381 | 0 |
| Library Technicians | 0.617647059 | 0.323529412 | 0.058823529 |
| Hoist and Winch Operators | 0.615384615 | 0.230769231 | 0.153846154 |
| Operators, and Tenders, Except Sawing | 0.615384615 | 0.269230769 | 0.115384615 |
| Workers | 0.615384615 | 0.307692308 | 0.076923077 |
| Legal Secretaries | 0.615384615 | 0.384615385 | 0 |
| Radio Mechanics | 0.615384615 | 0.384615385 | 0 |
| Manicurists and Pedicurists | 0.611111111 | 0.388888889 | 0 |
| Printing Press Operators | 0.608695652 | 0.173913043 | 0.217391304 |
| Dressing Room Attendants | 0.608695652 | 0.391304348 | 0 |
| Sailors and Marine Oilers | 0.607142857 | 0.357142857 | 0.035714286 |
| Bridge and Lock Tenders | 0.6 | 0.2 | 0.2 |
| Food Batchmakers | 0.6 | 0.36 | 0.04 |
| Locomotive Firers | 0.6 | 0.4 | 0 |
| Information Technicians | 0.6 | 0.4 | 0 |
| Paperhangers | 0.6 | 0.4 | 0 |
| Laundry and Dry-Cleaning Workers | 0.59375 | 0.40625 | 0 |
| Setters, Operators, and Tenders, Metal and Plastic | 0.58974359 | 0.41025641 | 0 |
| Ship Engineers | 0.588235294 | 0.294117647 | 0.117647059 |
| Baggage Porters and Bellhops | 0.588235294 | 0.411764706 | 0 |
| Billing, Cost, and Rate Clerks | 0.588235294 | 0.411764706 | 0 |
| Drawing Out Machine Setters, Operators, and Tenders | 0.583333333 | 0.291666667 | 0.125 |
| Operators | 0.583333333 | 0.333333333 | 0.083333333 |
| Retail Salespersons | 0.583333333 | 0.333333333 | 0.083333333 |
| Farm Products | 0.583333333 | 0.416666667 | 0 |
| News and Street Vendors, and Related Workers | 0.583333333 | 0.416666667 | 0 |
| Nursery Workers | 0.583333333 | 0.416666667 | 0 |
| Security Guards | 0.583333333 | 0.416666667 | 0 |
| Slot Supervisors | 0.583333333 | 0.416666667 | 0 |

| Occupation | | | |
|---|---|---|---|
| Procurement Clerks | 0.578947368 | 0.368421053 | 0.052631579 |
| Maintenance Equipment | 0.576923077 | 0.423076923 | 0 |
| Travel Agents | 0.571428571 | 0.142857143 | 0.285714286 |
| Helpers--Extraction Workers | 0.571428571 | 0.214285714 | 0.214285714 |
| Office Clerks, General | 0.571428571 | 0.333333333 | 0.095238095 |
| Photogrammetrists | 0.571428571 | 0.357142857 | 0.071428571 |
| Assemblers | 0.571428571 | 0.357142857 | 0.071428571 |
| License Clerks | 0.571428571 | 0.380952381 | 0.047619048 |
| Equipment and Systems Inspectors, Except Aviation | 0.571428571 | 0.380952381 | 0.047619048 |
| Farmworkers and Laborers, Crop | 0.571428571 | 0.428571429 | 0 |
| Damage | 0.571428571 | 0.428571429 | 0 |
| Postal Service Mail Carriers | 0.571428571 | 0.428571429 | 0 |
| Conveyor Operators and Tenders | 0.565217391 | 0.434782609 | 0 |
| Related Repairers | 0.564102564 | 0.41025641 | 0.025641026 |
| Counter and Rental Clerks | 0.5625 | 0.375 | 0.0625 |
| Insurance Policy Processing Clerks | 0.5625 | 0.375 | 0.0625 |
| Energy Brokers | 0.5625 | 0.4375 | 0 |
| Operators, and Tenders | 0.555555556 | 0.277777778 | 0.166666667 |
| Kettle Operators and Tenders | 0.555555556 | 0.388888889 | 0.055555556 |
| Database Architects | 0.555555556 | 0.444444444 | 0 |
| Loan Interviewers and Clerks | 0.555555556 | 0.444444444 | 0 |
| Court Reporters | 0.545454545 | 0.454545455 | 0 |
| Machine Operators and Tenders | 0.541666667 | 0.416666667 | 0.041666667 |
| Cargo and Freight Agents | 0.541666667 | 0.458333333 | 0 |
| Riggers | 0.538461538 | 0.230769231 | 0.230769231 |
| Workers | 0.538461538 | 0.307692308 | 0.153846154 |
| Agricultural Technicians | 0.538461538 | 0.346153846 | 0.115384615 |
| Specialists | 0.538461538 | 0.423076923 | 0.038461538 |
| Station Operators | 0.538461538 | 0.461538462 | 0 |
| Operators, and Tenders | 0.533333333 | 0.333333333 | 0.133333333 |
| Pest Control Workers | 0.533333333 | 0.333333333 | 0.133333333 |
| Equipment Repairers | 0.533333333 | 0.4 | 0.066666667 |
| New Accounts Clerks | 0.533333333 | 0.466666667 | 0 |
| Government Programs | 0.529411765 | 0.470588235 | 0 |
| Processing Machine Operators | 0.527777778 | 0.444444444 | 0.027777778 |
| and Finishers | 0.526315789 | 0.263157895 | 0.210526316 |
| Fallers | 0.526315789 | 0.368421053 | 0.105263158 |
| Clerks | 0.526315789 | 0.368421053 | 0.105263158 |
| Husbandry and Animal Care Workers | 0.526315789 | 0.421052632 | 0.052631579 |
| Commodities | 0.526315789 | 0.421052632 | 0.052631579 |
| Tenders | 0.526315789 | 0.421052632 | 0.052631579 |

| | | | |
|---|---|---|---|
| Attendants | 0.526315789 | 0.473684211 | 0 |
| Machine Servicers and Repairers | 0.526315789 | 0.473684211 | 0 |
| Tank Car, Truck, and Ship Loaders | 0.526315789 | 0.473684211 | 0 |
| Workers | 0.523809524 | 0.333333333 | 0.142857143 |
| Taxi Drivers and Chauffeurs | 0.523809524 | 0.380952381 | 0.095238095 |
| Motion Picture Projectionists | 0.523809524 | 0.476190476 | 0 |
| Parking Enforcement Workers | 0.52173913 | 0.391304348 | 0.086956522 |
| Mapping Technicians | 0.52 | 0.4 | 0.08 |
| Operators, and Hostlers | 0.52 | 0.48 | 0 |
| Workers | 0.518518519 | 0.481481481 | 0 |
| Drivers | 0.516129032 | 0.451612903 | 0.032258065 |
| Nursing Assistants | 0.515151515 | 0.424242424 | 0.060606061 |
| Farm Labor Contractors | 0.5 | 0 | 0.5 |
| and Dragline Operators | 0.5 | 0.125 | 0.375 |
| Inspectors | 0.5 | 0.166666667 | 0.333333333 |
| Assessors | 0.5 | 0.2 | 0.3 |
| Barbers | 0.5 | 0.2 | 0.3 |
| Ophthalmic Medical Technicians | 0.5 | 0.25 | 0.25 |
| Butchers and Meat Cutters | 0.5 | 0.333333333 | 0.166666667 |
| Operators, and Tenders, Metal and Plastic | 0.5 | 0.35 | 0.15 |
| Technicians | 0.5 | 0.357142857 | 0.142857143 |
| Flight Attendants | 0.5 | 0.357142857 | 0.142857143 |
| Accountants | 0.5 | 0.375 | 0.125 |
| Helpers--Roofers | 0.5 | 0.388888889 | 0.111111111 |
| and Runners | 0.5 | 0.409090909 | 0.090909091 |
| Meter Readers, Utilities | 0.5 | 0.416666667 | 0.083333333 |
| Telemarketers | 0.5 | 0.416666667 | 0.083333333 |
| Model Makers, Metal and Plastic | 0.5 | 0.428571429 | 0.071428571 |
| Multimedia Artists and Animators | 0.5 | 0.428571429 | 0.071428571 |
| Fabricators | 0.5 | 0.4375 | 0.0625 |
| Gaming Dealers | 0.5 | 0.45 | 0.05 |
| Machine Setters, Operators, and Tenders | 0.5 | 0.45 | 0.05 |
| Traders | 0.5 | 0.454545455 | 0.045454545 |
| Orderlies | 0.5 | 0.458333333 | 0.041666667 |
| Brokerage Clerks | 0.5 | 0.5 | 0 |
| Carpet Installers | 0.5 | 0.5 | 0 |
| Machine Setters, Operators, and Tenders | 0.5 | 0.5 | 0 |
| Dredge Operators | 0.5 | 0.5 | 0 |
| Driver/Sales Workers | 0.5 | 0.5 | 0 |
| Electronic Drafters | 0.5 | 0.5 | 0 |

| Occupation | Col1 | Col2 | Col3 |
|---|---|---|---|
| Setters, Operators, and Tenders, Metal and Plastic | 0.5 | 0.5 | 0 |
| Floor Sanders and Finishers | 0.5 | 0.5 | 0 |
| Products | 0.5 | 0.5 | 0 |
| Helpers--Electricians | 0.5 | 0.5 | 0 |
| Log Graders and Scalers | 0.5 | 0.5 | 0 |
| Maintenance Workers, Machinery | 0.5 | 0.5 | 0 |
| Patternmakers, Wood | 0.5 | 0.5 | 0 |
| Pourers and Casters, Metal | 0.5 | 0.5 | 0 |
| Radiation Therapists | 0.5 | 0.5 | 0 |
| Surgical Technologists | 0.5 | 0.5 | 0 |
| Repairers, Except Mechanical Door | 0.487179487 | 0.487179487 | 0.025641026 |
| Photographers | 0.482758621 | 0.448275862 | 0.068965517 |
| Office Machine Repairers | 0.48 | 0.48 | 0.04 |
| Technologists | 0.47826087 | 0.47826087 | 0.043478261 |
| Vocational Nurses | 0.47826087 | 0.52173913 | 0 |
| Radiologic Technicians | 0.47826087 | 0.52173913 | 0 |
| Animal Breeders | 0.476190476 | 0.428571429 | 0.095238095 |
| Forensic Science Technicians | 0.476190476 | 0.523809524 | 0 |
| Surveying Technicians | 0.473684211 | 0.263157895 | 0.263157895 |
| Gaming Managers | 0.473684211 | 0.421052632 | 0.105263158 |
| Gas Plant Operators | 0.473684211 | 0.526315789 | 0 |
| Graduate Teaching Assistants | 0.473684211 | 0.526315789 | 0 |
| Physical Therapist Aides | 0.473684211 | 0.526315789 | 0 |
| Software | 0.473684211 | 0.526315789 | 0 |
| Technicians | 0.470588235 | 0 | 0.529411765 |
| Interpreters and Translators | 0.470588235 | 0.470588235 | 0.058823529 |
| Operators and Tenders | 0.470588235 | 0.529411765 | 0 |
| Samplers, and Weighers | 0.46875 | 0.40625 | 0.125 |
| Technical Directors/Managers | 0.466666667 | 0.266666667 | 0.266666667 |
| Craft Artists | 0.466666667 | 0.333333333 | 0.2 |
| Survey Researchers | 0.466666667 | 0.333333333 | 0.2 |
| Technicians | 0.466666667 | 0.4 | 0.133333333 |
| Client | 0.466666667 | 0.466666667 | 0.066666667 |
| Food Science Technicians | 0.466666667 | 0.466666667 | 0.066666667 |
| Workers | 0.466666667 | 0.5 | 0.033333333 |
| Medical Transcriptionists | 0.466666667 | 0.533333333 | 0 |
| Roofers | 0.464285714 | 0.428571429 | 0.107142857 |
| Drywall and Ceiling Tile Installers | 0.461538462 | 0.230769231 | 0.307692308 |
| Sewing Machine Operators | 0.461538462 | 0.461538462 | 0.076923077 |
| Bus Drivers, Transit and Intercity | 0.461538462 | 0.538461538 | 0 |
| and Ambulance | 0.461538462 | 0.538461538 | 0 |

| Occupation | Col1 | Col2 | Col3 |
|---|---|---|---|
| and Trimmers | 0.461538462 | 0.538461538 | 0 |
| Technicians | 0.461538462 | 0.538461538 | 0 |
| Automotive Master Mechanics | 0.458333333 | 0.416666667 | 0.125 |
| Nurse Anesthetists | 0.458333333 | 0.458333333 | 0.083333333 |
| Construction Laborers | 0.454545455 | 0.393939394 | 0.151515152 |
| Sewers | 0.454545455 | 0.5 | 0.045454545 |
| Aquacultural Animals | 0.454545455 | 0.545454545 | 0 |
| Warehouse, or Storage Yard | 0.454545455 | 0.545454545 | 0 |
| Technicians | 0.451612903 | 0.451612903 | 0.096774194 |
| Jewelers | 0.45 | 0.35 | 0.2 |
| Carpenters | 0.45 | 0.45 | 0.1 |
| Cooks, Restaurant | 0.45 | 0.45 | 0.1 |
| Clerks | 0.45 | 0.55 | 0 |
| Rail Car Repairers | 0.45 | 0.55 | 0 |
| Construction Equipment Operators | 0.448275862 | 0.413793103 | 0.137931034 |
| Repairers | 0.444444444 | 0.407407407 | 0.148148148 |
| Desktop Publishers | 0.444444444 | 0.5 | 0.055555556 |
| Online Merchants | 0.441176471 | 0.529411765 | 0.029411765 |
| Repairers | 0.44 | 0.52 | 0.04 |
| Technicians | 0.44 | 0.52 | 0.04 |
| Mechanical Door Repairers | 0.44 | 0.52 | 0.04 |
| Equipment | 0.44 | 0.56 | 0 |
| Teacher Assistants | 0.4375 | 0.375 | 0.1875 |
| Biomass Plant Technicians | 0.4375 | 0.5 | 0.0625 |
| and Analysts | 0.434782609 | 0.347826087 | 0.217391304 |
| Technologists | 0.434782609 | 0.565217391 | 0 |
| Assistants | 0.434782609 | 0.565217391 | 0 |
| Traffic Technicians | 0.434782609 | 0.565217391 | 0 |
| Cytogenetic Technologists | 0.433333333 | 0.433333333 | 0.133333333 |
| Earth Drillers, Except Oil and Gas | 0.433333333 | 0.5 | 0.066666667 |
| Food Service Managers | 0.428571429 | 0.285714286 | 0.285714286 |
| Forest and Conservation Workers | 0.428571429 | 0.476190476 | 0.095238095 |
| Nonfarm Animal Caretakers | 0.428571429 | 0.476190476 | 0.095238095 |
| Drivers | 0.428571429 | 0.5 | 0.071428571 |
| Model Makers, Wood | 0.428571429 | 0.5 | 0.071428571 |
| Serving Workers, Including Fast Food | 0.428571429 | 0.523809524 | 0.047619048 |
| Setters, Operators, and Tenders | 0.428571429 | 0.535714286 | 0.035714286 |
| Credit Checkers | 0.428571429 | 0.571428571 | 0 |
| Wood, and Hard Tiles | 0.428571429 | 0.571428571 | 0 |
| Insurance Underwriters | 0.428571429 | 0.571428571 | 0 |
| Layout Workers, Metal and Plastic | 0.428571429 | 0.571428571 | 0 |
| Molding and Casting Workers | 0.428571429 | 0.571428571 | 0 |

| Occupation | | | |
|---|---|---|---|
| Officers | 0.428571429 | 0.571428571 | 0 |
| Workers | 0.428571429 | 0.571428571 | 0 |
| Medical Equipment Repairers | 0.421052632 | 0.473684211 | 0.105263158 |
| Biomedical Engineers | 0.421052632 | 0.526315789 | 0.052631579 |
| Biofuels Processing Technicians | 0.421052632 | 0.578947368 | 0 |
| Agricultural Crop and Horticultural Workers | 0.416666667 | 0.375 | 0.208333333 |
| Geographers | 0.416666667 | 0.416666667 | 0.166666667 |
| Subway and Streetcar Operators | 0.416666667 | 0.416666667 | 0.166666667 |
| Laborers, and Material Movers, Hand | 0.416666667 | 0.458333333 | 0.125 |
| Flight Attendants | 0.416666667 | 0.458333333 | 0.125 |
| Promoters | 0.416666667 | 0.541666667 | 0.041666667 |
| Paralegals and Legal Assistants | 0.416666667 | 0.583333333 | 0 |
| Phlebotomists | 0.416666667 | 0.583333333 | 0 |
| Ship and Boat Captains | 0.416666667 | 0.583333333 | 0 |
| Technicians | 0.413793103 | 0.448275862 | 0.137931034 |
| Industrial Engineering Technicians | 0.411764706 | 0.470588235 | 0.117647059 |
| Dental Laboratory Technicians | 0.411764706 | 0.529411765 | 0.058823529 |
| Psychiatric Aides | 0.411764706 | 0.529411765 | 0.058823529 |
| Setters, Operators, and Tenders, Metal and Plastic | 0.411764706 | 0.588235294 | 0 |
| Motorboat Operators | 0.411764706 | 0.588235294 | 0 |
| Upholsterers | 0.409090909 | 0.272727273 | 0.318181818 |
| Gas | 0.409090909 | 0.5 | 0.090909091 |
| Pipe Cleaners | 0.409090909 | 0.590909091 | 0 |
| Laboratory Animal Caretakers | 0.407407407 | 0.518518519 | 0.074074074 |
| Precious Metal Workers | 0.40625 | 0.59375 | 0 |
| Diesel Engine Specialists | 0.4 | 0.44 | 0.16 |
| Bartenders | 0.4 | 0.45 | 0.15 |
| Diagnostic Medical Sonographers | 0.4 | 0.45 | 0.15 |
| Machine Tool Programmers, Metal and Plastic | 0.4 | 0.466666667 | 0.133333333 |
| Industrial Machinery Mechanics | 0.4 | 0.466666667 | 0.133333333 |
| Hearing Aid Specialists | 0.4 | 0.5 | 0.1 |
| and Wall | 0.4 | 0.5 | 0.1 |
| Pipe Fitters and Steamfitters | 0.4 | 0.5 | 0.1 |
| Setters, Operators, and Tenders, Metal and Plastic | 0.4 | 0.533333333 | 0.066666667 |
| Auditors | 0.4 | 0.55 | 0.05 |
| Operators | 0.4 | 0.56 | 0.04 |
| Web Administrators | 0.4 | 0.571428571 | 0.028571429 |
| Concierges | 0.4 | 0.6 | 0 |

| Occupation | | | |
|---|---|---|---|
| Medical Appliance Technicians | 0.4 | 0.6 | 0 |
| Occupational Therapy Aides | 0.4 | 0.6 | 0 |
| Slaughterers and Meat Packers | 0.4 | 0.6 | 0 |
| Software Developers, Applications | 0.4 | 0.6 | 0 |
| Tile and Marble Setters | 0.4 | 0.6 | 0 |
| and Tenders | 0.391304348 | 0.434782609 | 0.173913043 |
| Ticket Takers | 0.391304348 | 0.47826087 | 0.130434783 |
| Specialists | 0.391304348 | 0.565217391 | 0.043478261 |
| Municipal Clerks | 0.391304348 | 0.565217391 | 0.043478261 |
| Buffing Machine Tool Setters, Operators, and Tenders, Metal | 0.388888889 | 0.444444444 | 0.166666667 |
| Librarians | 0.388888889 | 0.444444444 | 0.166666667 |
| Retail Sales Workers | 0.388888889 | 0.5 | 0.111111111 |
| Technicians | 0.388888889 | 0.5 | 0.111111111 |
| Collections Specialists | 0.388888889 | 0.555555556 | 0.055555556 |
| Helpers--Carpenters | 0.388888889 | 0.611111111 | 0 |
| Except Payroll and Timekeeping | 0.388888889 | 0.611111111 | 0 |
| Home Appliance Repairers | 0.387096774 | 0.580645161 | 0.032258065 |
| Fabric and Apparel Patternmakers | 0.384615385 | 0.461538462 | 0.153846154 |
| Food Scientists and Technologists | 0.384615385 | 0.461538462 | 0.153846154 |
| Tuners | 0.384615385 | 0.512820513 | 0.102564103 |
| Funeral Attendants | 0.384615385 | 0.538461538 | 0.076923077 |
| Wellhead Pumpers | 0.384615385 | 0.538461538 | 0.076923077 |
| Wind Turbine Service Technicians | 0.384615385 | 0.538461538 | 0.076923077 |
| Chefs and Head Cooks | 0.380952381 | 0.428571429 | 0.19047619 |
| Technologists | 0.380952381 | 0.428571429 | 0.19047619 |
| Device Specialists | 0.380952381 | 0.428571429 | 0.19047619 |
| Electricians | 0.380952381 | 0.476190476 | 0.142857143 |
| Sales Workers | 0.380952381 | 0.523809524 | 0.095238095 |
| Broadcast Technicians | 0.379310345 | 0.517241379 | 0.103448276 |
| Appraisers, Real Estate | 0.375 | 0.5625 | 0.0625 |
| Aquacultural Workers | 0.375 | 0.5625 | 0.0625 |
| Geodetic Surveyors | 0.375 | 0.5625 | 0.0625 |
| Installers and Repairers, Except Line Installers | 0.375 | 0.575 | 0.05 |
| Floral Designers | 0.375 | 0.625 | 0 |
| Proofreaders and Copy Markers | 0.375 | 0.625 | 0 |
| Related Materials | 0.37037037 | 0.592592593 | 0.037037037 |
| Actors | 0.368421053 | 0.315789474 | 0.315789474 |
| Animal Trainers | 0.368421053 | 0.421052632 | 0.210526316 |
| Wholesale, Retail, and Farm | 0.368421053 | 0.473684211 | 0.157894737 |
| Maintenance | 0.368421053 | 0.526315789 | 0.105263158 |
| Technicians | 0.368421053 | 0.578947368 | 0.052631579 |

| Occupation | | | |
|---|---|---|---|
| Pathologists | 0.368421053 | 0.578947368 | 0.052631579 |
| Technicians | 0.368421053 | 0.631578947 | 0 |
| Assistants | 0.363636364 | 0.454545455 | 0.181818182 |
| Operations Technicians | 0.363636364 | 0.545454545 | 0.090909091 |
| Lounge, and Coffee Shop | 0.363636364 | 0.636363636 | 0 |
| Yardmasters | 0.363636364 | 0.636363636 | 0 |
| Tax Preparers | 0.363636364 | 0.636363636 | 0 |
| Transit and Railroad Police | 0.363636364 | 0.636363636 | 0 |
| Embalmers | 0.36 | 0.36 | 0.28 |
| Human Resources Specialists | 0.36 | 0.56 | 0.08 |
| Sound Engineering Technicians | 0.357142857 | 0.428571429 | 0.214285714 |
| Urologists | 0.357142857 | 0.5 | 0.142857143 |
| Mechanical Drafters | 0.357142857 | 0.571428571 | 0.071428571 |
| Pipelayers | 0.357142857 | 0.571428571 | 0.071428571 |
| Pumpers | 0.357142857 | 0.571428571 | 0.071428571 |
| Watch Repairers | 0.357142857 | 0.571428571 | 0.071428571 |
| Administrative Services Managers | 0.357142857 | 0.642857143 | 0 |
| Biofuels Production Managers | 0.357142857 | 0.642857143 | 0 |
| Logistics Analysts | 0.35483871 | 0.516129032 | 0.129032258 |
| Police Detectives | 0.35483871 | 0.64516129 | 0 |
| Investigators | 0.352941176 | 0.470588235 | 0.176470588 |
| General and Operations Managers | 0.352941176 | 0.647058824 | 0 |
| Tool and Die Makers | 0.352941176 | 0.647058824 | 0 |
| Potters, Manufacturing | 0.35 | 0.5 | 0.15 |
| Except Farm Products | 0.35 | 0.5 | 0.15 |
| Fashion Designers | 0.35 | 0.6 | 0.05 |
| Administrators | 0.35 | 0.6 | 0.05 |
| Pharmacists | 0.35 | 0.65 | 0 |
| Court Clerks | 0.347826087 | 0.260869565 | 0.391304348 |
| and Repairers | 0.347826087 | 0.52173913 | 0.130434783 |
| Technologists | 0.347826087 | 0.565217391 | 0.086956522 |
| Transportation Security Screeners | 0.346153846 | 0.461538462 | 0.192307692 |
| Childcare Workers | 0.346153846 | 0.538461538 | 0.115384615 |
| Housekeeping and Janitorial | 0.346153846 | 0.576923077 | 0.076923077 |
| Automotive Specialty Technicians | 0.346153846 | 0.615384615 | 0.038461538 |
| Video, and Motion Picture | 0.346153846 | 0.615384615 | 0.038461538 |
| Specialists | 0.346153846 | 0.615384615 | 0.038461538 |
| Solar Photovoltaic Installers | 0.346153846 | 0.653846154 | 0 |
| Revenue Agents | 0.346153846 | 0.653846154 | 0 |
| Critical Care Nurses | 0.344827586 | 0.413793103 | 0.24137931 |
| Handling Experts, and Blasters | 0.344827586 | 0.586206897 | 0.068965517 |
| Customer Service Representatives | 0.333333333 | 0.266666667 | 0.4 |
| Home Health Aides | 0.333333333 | 0.266666667 | 0.4 |

| Occupation | | | |
|---|---|---|---|
| Municipal Firefighters | 0.333333333 | 0.407407407 | 0.259259259 |
| Assemblers | 0.333333333 | 0.416666667 | 0.25 |
| Installers | 0.333333333 | 0.476190476 | 0.19047619 |
| and Repairers, Motor Vehicles | 0.333333333 | 0.5 | 0.166666667 |
| Clinical Data Managers | 0.333333333 | 0.523809524 | 0.142857143 |
| Opticians, Dispensing | 0.333333333 | 0.523809524 | 0.142857143 |
| Radio and Television Announcers | 0.333333333 | 0.523809524 | 0.142857143 |
| Bill and Account Collectors | 0.333333333 | 0.533333333 | 0.133333333 |
| Civil Drafters | 0.333333333 | 0.533333333 | 0.133333333 |
| Power Plant Operators | 0.333333333 | 0.541666667 | 0.125 |
| Logging Equipment Operators | 0.333333333 | 0.555555556 | 0.111111111 |
| Funeral Directors | 0.333333333 | 0.571428571 | 0.095238095 |
| Technicians | 0.333333333 | 0.571428571 | 0.095238095 |
| City and Regional Planning Aides | 0.333333333 | 0.583333333 | 0.083333333 |
| Information Security Analysts | 0.333333333 | 0.583333333 | 0.083333333 |
| Lodging Managers | 0.333333333 | 0.583333333 | 0.083333333 |
| General | 0.333333333 | 0.6 | 0.066666667 |
| Dispatchers | 0.333333333 | 0.6 | 0.066666667 |
| Radio Operators | 0.333333333 | 0.6 | 0.066666667 |
| Rough Carpenters | 0.333333333 | 0.6 | 0.066666667 |
| Soil and Water Conservationists | 0.333333333 | 0.606060606 | 0.060606061 |
| Loss Prevention Managers | 0.333333333 | 0.62962963 | 0.037037037 |
| Bicycle Repairers | 0.333333333 | 0.666666667 | 0 |
| Costume Attendants | 0.333333333 | 0.666666667 | 0 |
| Blockmasons, Stonemasons, and Tile and Marble Setters | 0.333333333 | 0.666666667 | 0 |
| System Operators | 0.333333333 | 0.666666667 | 0 |
| Plasterers and Stucco Masons | 0.333333333 | 0.666666667 | 0 |
| Superintendents | 0.333333333 | 0.666666667 | 0 |
| Spa Managers | 0.333333333 | 0.666666667 | 0 |
| Managers | 0.333333333 | 0.666666667 | 0 |
| Sharpeners | 0.333333333 | 0.666666667 | 0 |
| Freight Forwarders | 0.322580645 | 0.612903226 | 0.064516129 |
| Conservators | 0.32 | 0.44 | 0.24 |
| Window Trimmers | 0.32 | 0.6 | 0.08 |
| Protection Technicians, Including Health | 0.32 | 0.68 | 0 |
| Respiratory Therapists | 0.318181818 | 0.045454545 | 0.636363636 |
| Transportation and Material-Moving Machine and Vehicle | 0.318181818 | 0.454545455 | 0.227272727 |
| Cosmetologists | 0.318181818 | 0.5 | 0.181818182 |
| Remote Sensing Technicians | 0.318181818 | 0.590909091 | 0.090909091 |
| Audiologists | 0.318181818 | 0.636363636 | 0.045454545 |

| Occupation | | | |
|---|---|---|---|
| Executive Administrative | 0.318181818 | 0.636363636 | 0.045454545 |
| Furniture Finishers | 0.318181818 | 0.681818182 | 0 |
| Production and Operating Workers | 0.315789474 | 0.210526316 | 0.473684211 |
| Technicians | 0.315789474 | 0.578947368 | 0.105263158 |
| Nuclear Monitoring Technicians | 0.315789474 | 0.631578947 | 0.052631579 |
| Operators | 0.315789474 | 0.684210526 | 0 |
| Chemical Technicians | 0.3125 | 0.5 | 0.1875 |
| Repairers, Powerhouse, Substation, and Relay | 0.3125 | 0.5 | 0.1875 |
| Installers | 0.3125 | 0.5625 | 0.125 |
| Etchers and Engravers | 0.3125 | 0.65625 | 0.03125 |
| Biological Technicians | 0.3125 | 0.6875 | 0 |
| Casualty Insurance | 0.3125 | 0.6875 | 0 |
| Maintenance, and Repair Workers | 0.3125 | 0.6875 | 0 |
| Chemists | 0.307692308 | 0.461538462 | 0.230769231 |
| Epidemiologists | 0.307692308 | 0.461538462 | 0.230769231 |
| Terrazzo Workers and Finishers | 0.307692308 | 0.576923077 | 0.115384615 |
| Assessors | 0.307692308 | 0.615384615 | 0.076923077 |
| Commercial Pilots | 0.307692308 | 0.653846154 | 0.038461538 |
| Archivists | 0.307692308 | 0.692307692 | 0 |
| Loan Officers | 0.304347826 | 0.565217391 | 0.130434783 |
| Radiologic Technologists | 0.304347826 | 0.608695652 | 0.086956522 |
| Recycling Coordinators | 0.304347826 | 0.652173913 | 0.043478261 |
| Coroners | 0.3 | 0.4 | 0.3 |
| Materials Engineers | 0.3 | 0.55 | 0.15 |
| Advertising Sales Agents | 0.3 | 0.6 | 0.1 |
| Naturopathic Physicians | 0.3 | 0.6 | 0.1 |
| Rigging, and Systems Assemblers | 0.3 | 0.7 | 0 |
| Credit Analysts | 0.3 | 0.7 | 0 |
| and Manufacturing, Except Technical and Scientific Products | 0.3 | 0.7 | 0 |
| Graphic Designers | 0.294117647 | 0.411764706 | 0.294117647 |
| Expediting Clerks | 0.294117647 | 0.588235294 | 0.117647059 |
| Business Intelligence Analysts | 0.294117647 | 0.647058824 | 0.058823529 |
| Systems Managers | 0.294117647 | 0.647058824 | 0.058823529 |
| Dental Hygienists | 0.294117647 | 0.647058824 | 0.058823529 |
| Technologists | 0.294117647 | 0.647058824 | 0.058823529 |
| Architectural Drafters | 0.294117647 | 0.705882353 | 0 |
| Civil Engineers | 0.294117647 | 0.705882353 | 0 |
| Financial Examiners | 0.294117647 | 0.705882353 | 0 |
| Service Workers | 0.294117647 | 0.705882353 | 0 |
| System Technicians | 0.294117647 | 0.705882353 | 0 |
| Travel Guides | 0.294117647 | 0.705882353 | 0 |

| Occupation | | | |
|---|---|---|---|
| Geneticists | 0.291666667 | 0.541666667 | 0.166666667 |
| Technicians | 0.289473684 | 0.710526316 | 0 |
| Gaming Supervisors | 0.285714286 | 0.19047619 | 0.523809524 |
| Energy Auditors | 0.285714286 | 0.285714286 | 0.428571429 |
| Energy Engineers | 0.285714286 | 0.428571429 | 0.285714286 |
| Loan Counselors | 0.285714286 | 0.5 | 0.214285714 |
| Prevention Supervisors | 0.285714286 | 0.535714286 | 0.178571429 |
| Geophysical Data Technicians | 0.285714286 | 0.571428571 | 0.142857143 |
| Registered Nurses | 0.285714286 | 0.607142857 | 0.107142857 |
| Agricultural Engineers | 0.285714286 | 0.642857143 | 0.071428571 |
| Prevention Specialists | 0.285714286 | 0.642857143 | 0.071428571 |
| Other Small Engine Mechanics | 0.285714286 | 0.642857143 | 0.071428571 |
| Hydroelectric Plant Technicians | 0.285714286 | 0.666666667 | 0.047619048 |
| Engineers and Testers | 0.285714286 | 0.678571429 | 0.035714286 |
| Mobile Home Installers | 0.285714286 | 0.714285714 | 0 |
| Patternmakers, Metal and Plastic | 0.285714286 | 0.714285714 | 0 |
| Postsecondary | 0.28 | 0.4 | 0.32 |
| Construction Managers | 0.28 | 0.56 | 0.16 |
| Sales Engineers | 0.28 | 0.6 | 0.12 |
| and Manufacturing, Technical and Scientific Products | 0.277777778 | 0.527777778 | 0.194444444 |
| Mining and Geological Engineers, Including Mining Safety Engineers | 0.277777778 | 0.611111111 | 0.111111111 |
| Dispatchers | 0.277777778 | 0.611111111 | 0.111111111 |
| Preschool | 0.277777778 | 0.611111111 | 0.111111111 |
| Computer Hardware Engineers | 0.277777778 | 0.666666667 | 0.055555556 |
| Physical Therapist Assistants | 0.277777778 | 0.666666667 | 0.055555556 |
| Acupuncturists | 0.277777778 | 0.722222222 | 0 |
| Ophthalmic Medical Technologists | 0.275862069 | 0.620689655 | 0.103448276 |
| Team Assemblers | 0.272727273 | 0.545454545 | 0.181818182 |
| Music Composers and Arrangers | 0.272727273 | 0.590909091 | 0.136363636 |
| Computer | 0.272727273 | 0.681818182 | 0.045454545 |
| Mechanics, Installers, and | 0.272727273 | 0.727272727 | 0 |
| Inspectors | 0.272727273 | 0.727272727 | 0 |
| Management Analysts | 0.272727273 | 0.727272727 | 0 |
| Mates- Ship, Boat, and Barge | 0.272727273 | 0.727272727 | 0 |
| Technicians | 0.269230769 | 0.615384615 | 0.115384615 |
| Correctional Officers and Jailers | 0.269230769 | 0.653846154 | 0.076923077 |
| Preparation and Serving Workers | 0.269230769 | 0.653846154 | 0.076923077 |
| Engineers | 0.269230769 | 0.692307692 | 0.038461538 |
| Finishers | 0.269230769 | 0.692307692 | 0.038461538 |
| Inspectors | 0.269230769 | 0.692307692 | 0.038461538 |
| Zoologists and Wildlife Biologists | 0.266666667 | 0.2 | 0.533333333 |

| Occupation | | | |
|---|---|---|---|
| Installers and Repairers, Transportation Equipment | 0.266666667 | 0.4 | 0.333333333 |
| Recreational Protective Service Workers | 0.266666667 | 0.4 | 0.333333333 |
| Orthotists and Prosthetists | 0.266666667 | 0.533333333 | 0.2 |
| Locomotive Engineers | 0.266666667 | 0.6 | 0.133333333 |
| Computer Programmers | 0.266666667 | 0.666666667 | 0.066666667 |
| Setters, Operators, and Tenders, Metal and Plastic | 0.266666667 | 0.666666667 | 0.066666667 |
| Pipefitters, and Steamfitters | 0.266666667 | 0.733333333 | 0 |
| Photonics Technicians | 0.266666667 | 0.733333333 | 0 |
| Sports Officials | 0.266666667 | 0.733333333 | 0 |
| Real Estate Brokers | 0.263157895 | 0.157894737 | 0.578947368 |
| Tour Guides and Escorts | 0.263157895 | 0.421052632 | 0.315789474 |
| Aquacultural Managers | 0.263157895 | 0.473684211 | 0.263157895 |
| Bioinformatics Technicians | 0.263157895 | 0.578947368 | 0.157894737 |
| Judicial Law Clerks | 0.263157895 | 0.684210526 | 0.052631579 |
| and Mining | 0.263157895 | 0.684210526 | 0.052631579 |
| Plumbers | 0.260869565 | 0.391304348 | 0.347826087 |
| Fitness and Wellness Coordinators | 0.260869565 | 0.565217391 | 0.173913043 |
| Precision Agriculture Technicians | 0.260869565 | 0.608695652 | 0.130434783 |
| Marine Engineers | 0.260869565 | 0.652173913 | 0.086956522 |
| Mechatronics Engineers | 0.260869565 | 0.652173913 | 0.086956522 |
| Distance Learning Coordinators | 0.260869565 | 0.695652174 | 0.043478261 |
| Credit Counselors | 0.260869565 | 0.739130435 | 0 |
| Farm and Ranch Managers | 0.259259259 | 0.62962963 | 0.111111111 |
| Business Teachers, Postsecondary | 0.25 | 0.25 | 0.5 |
| Park Naturalists | 0.25 | 0.25 | 0.5 |
| Manufacturing | 0.25 | 0.25 | 0.5 |
| Teachers, Postsecondary | 0.25 | 0.375 | 0.375 |
| Managers | 0.25 | 0.4 | 0.35 |
| Teachers, Postsecondary | 0.25 | 0.416666667 | 0.333333333 |
| Postsecondary | 0.25 | 0.416666667 | 0.333333333 |
| Physics Teachers, Postsecondary | 0.25 | 0.458333333 | 0.291666667 |
| Broadcast News Analysts | 0.25 | 0.5 | 0.25 |
| Shampooers | 0.25 | 0.5 | 0.25 |
| Managers | 0.25 | 0.5 | 0.25 |
| Elevator Installers and Repairers | 0.25 | 0.55 | 0.2 |
| Archeologists | 0.25 | 0.5625 | 0.1875 |
| Crossing Guards | 0.25 | 0.583333333 | 0.166666667 |
| Real Estate Sales Agents | 0.25 | 0.59375 | 0.15625 |
| Landscaping, Lawn Service, and Groundskeeping Workers | 0.25 | 0.607142857 | 0.142857143 |

| Occupation | | | |
|---|---|---|---|
| Mechanical Engineers | 0.25 | 0.607142857 | 0.142857143 |
| Technologists | 0.25 | 0.625 | 0.125 |
| Repairers, Commercial and Industrial Equipment | 0.25 | 0.65 | 0.1 |
| Police Patrol Officers | 0.25 | 0.65 | 0.1 |
| Fundraisers | 0.25 | 0.678571429 | 0.071428571 |
| Animal Control Workers | 0.25 | 0.6875 | 0.0625 |
| Specialists | 0.25 | 0.6875 | 0.0625 |
| Loan | 0.25 | 0.6875 | 0.0625 |
| Robotics Engineers | 0.25 | 0.708333333 | 0.041666667 |
| Technicians | 0.25 | 0.75 | 0 |
| Operators | 0.25 | 0.75 | 0 |
| Except Brickmasons | 0.25 | 0.75 | 0 |
| Web Developers | 0.243243243 | 0.702702703 | 0.054054054 |
| Clinical Research Coordinators | 0.242424242 | 0.696969697 | 0.060606061 |
| History Teachers, Postsecondary | 0.24 | 0.44 | 0.32 |
| Postsecondary | 0.24 | 0.52 | 0.24 |
| Commercial Divers | 0.24 | 0.72 | 0.04 |
| Exercise Physiologists | 0.24 | 0.72 | 0.04 |
| Department | 0.238095238 | 0.714285714 | 0.047619048 |
| Planners | 0.238095238 | 0.761904762 | 0 |
| Boilermakers | 0.235294118 | 0.411764706 | 0.352941176 |
| Purchasing Managers | 0.235294118 | 0.588235294 | 0.176470588 |
| Technicians | 0.235294118 | 0.588235294 | 0.176470588 |
| Nuclear Medicine Technologists | 0.235294118 | 0.705882353 | 0.058823529 |
| Occupational Therapists | 0.235294118 | 0.705882353 | 0.058823529 |
| Special Education | 0.235294118 | 0.764705882 | 0 |
| Residential Advisors | 0.233333333 | 0.7 | 0.066666667 |
| Sewers, Hand | 0.230769231 | 0.769230769 | 0 |
| Special and Career/Technical Education | 0.228571429 | 0.742857143 | 0.028571429 |
| Postsecondary | 0.227272727 | 0.363636364 | 0.409090909 |
| Postsecondary | 0.227272727 | 0.409090909 | 0.363636364 |
| Enforcement Teachers, | 0.227272727 | 0.409090909 | 0.363636364 |
| Postsecondary | 0.227272727 | 0.409090909 | 0.363636364 |
| Gem and Diamond Workers | 0.227272727 | 0.409090909 | 0.363636364 |
| Postsecondary | 0.227272727 | 0.409090909 | 0.363636364 |
| Postsecondary | 0.227272727 | 0.409090909 | 0.363636364 |
| Occupational Therapy Assistants | 0.227272727 | 0.409090909 | 0.363636364 |
| Postsecondary | 0.227272727 | 0.409090909 | 0.363636364 |
| Forest Firefighters | 0.227272727 | 0.681818182 | 0.090909091 |
| Agents | 0.227272727 | 0.772727273 | 0 |
| Transportation Planners | 0.227272727 | 0.772727273 | 0 |

| Occupation | | | |
|---|---|---|---|
| Middle School | 0.225 | 0.725 | 0.05 |
| Skincare Specialists | 0.222222222 | 0.388888889 | 0.388888889 |
| Postsecondary | 0.222222222 | 0.518518519 | 0.259259259 |
| Attendants, Except Emergency Medical Technicians | 0.222222222 | 0.666666667 | 0.111111111 |
| Pilots, Ship | 0.222222222 | 0.666666667 | 0.111111111 |
| Acute Care Nurses | 0.222222222 | 0.703703704 | 0.074074074 |
| Financial Analysts | 0.222222222 | 0.722222222 | 0.055555556 |
| Intelligence Analysts | 0.222222222 | 0.777777778 | 0 |
| Community Association Managers | 0.222222222 | 0.777777778 | 0 |
| Statisticians | 0.222222222 | 0.777777778 | 0 |
| Except Special and Career/Technical Education | 0.21875 | 0.71875 | 0.0625 |
| Postsecondary | 0.217391304 | 0.391304348 | 0.391304348 |
| Teachers, Postsecondary | 0.217391304 | 0.434782609 | 0.347826087 |
| Postsecondary | 0.217391304 | 0.434782609 | 0.347826087 |
| Postsecondary | 0.217391304 | 0.434782609 | 0.347826087 |
| Postsecondary | 0.217391304 | 0.434782609 | 0.347826087 |
| Postsecondary | 0.217391304 | 0.434782609 | 0.347826087 |
| Sociology Teachers, Postsecondary | 0.217391304 | 0.434782609 | 0.347826087 |
| Law Teachers, Postsecondary | 0.217391304 | 0.47826087 | 0.304347826 |
| Veterinarians | 0.217391304 | 0.52173913 | 0.260869565 |
| Robotics Technicians | 0.217391304 | 0.608695652 | 0.173913043 |
| Film and Video Editors | 0.217391304 | 0.652173913 | 0.130434783 |
| Air Traffic Controllers | 0.217391304 | 0.739130435 | 0.043478261 |
| Customs Brokers | 0.217391304 | 0.739130435 | 0.043478261 |
| Surveyors | 0.217391304 | 0.782608696 | 0 |
| Industrial Production Managers | 0.214285714 | 0.571428571 | 0.214285714 |
| Environmental Engineers | 0.214285714 | 0.607142857 | 0.178571429 |
| Service Technicians | 0.214285714 | 0.785714286 | 0 |
| Insurance Sales Agents | 0.210526316 | 0.631578947 | 0.157894737 |
| Operators, and Tenders, Metal and Plastic | 0.210526316 | 0.631578947 | 0.157894737 |
| Internists, General | 0.210526316 | 0.736842105 | 0.052631579 |
| Postsecondary | 0.208333333 | 0.5 | 0.291666667 |
| Postsecondary | 0.208333333 | 0.5 | 0.291666667 |
| Manufacturing Engineers | 0.208333333 | 0.583333333 | 0.208333333 |
| Security Management Specialists | 0.208333333 | 0.625 | 0.166666667 |
| Nurses | 0.208333333 | 0.708333333 | 0.083333333 |
| Geothermal Technicians | 0.208333333 | 0.75 | 0.041666667 |
| and Administrative Support Workers | 0.206896552 | 0.551724138 | 0.24137931 |
| Machinists | 0.206896552 | 0.724137931 | 0.068965517 |

| Occupation | | | |
|---|---|---|---|
| Radiologists | 0.206896552 | 0.793103448 | 0 |
| Construction Trades and Extraction Workers | 0.2 | 0.266666667 | 0.533333333 |
| Midwives | 0.2 | 0.4 | 0.4 |
| Teachers, Postsecondary | 0.2 | 0.466666667 | 0.333333333 |
| Managers | 0.2 | 0.48 | 0.32 |
| Teachers, Postsecondary | 0.2 | 0.48 | 0.32 |
| Aviation Inspectors | 0.2 | 0.533333333 | 0.266666667 |
| Technical Writers | 0.2 | 0.533333333 | 0.266666667 |
| Derrick Operators, Oil and Gas | 0.2 | 0.6 | 0.2 |
| Nurse Midwives | 0.2 | 0.6 | 0.2 |
| Hydrologists | 0.2 | 0.64 | 0.16 |
| and Paramedics | 0.2 | 0.666666667 | 0.133333333 |
| and Detectives | 0.2 | 0.7 | 0.1 |
| Investment Fund Managers | 0.2 | 0.7 | 0.1 |
| Specialists | 0.2 | 0.7 | 0.1 |
| Family and General Practitioners | 0.2 | 0.733333333 | 0.066666667 |
| Materials Scientists | 0.2 | 0.733333333 | 0.066666667 |
| Physicists | 0.2 | 0.733333333 | 0.066666667 |
| Teachers | 0.2 | 0.733333333 | 0.066666667 |
| Managers | 0.2 | 0.733333333 | 0.066666667 |
| Nuclear Power Reactor Operators | 0.2 | 0.75 | 0.05 |
| Secondary School | 0.2 | 0.75 | 0.05 |
| Art Therapists | 0.2 | 0.8 | 0 |
| Gaming Investigators | 0.2 | 0.8 | 0 |
| Industrial Engineers | 0.2 | 0.8 | 0 |
| Sheriffs and Deputy Sheriffs | 0.2 | 0.8 | 0 |
| Teachers, Middle School | 0.193548387 | 0.741935484 | 0.064516129 |
| Technicians | 0.192307692 | 0.346153846 | 0.461538462 |
| Space Sciences Teachers, Postsecondary | 0.192307692 | 0.461538462 | 0.346153846 |
| Postsecondary | 0.192307692 | 0.538461538 | 0.269230769 |
| Mechanics and Installers | 0.192307692 | 0.730769231 | 0.076923077 |
| Performance | 0.19047619 | 0.476190476 | 0.333333333 |
| Business Continuity Planners | 0.19047619 | 0.571428571 | 0.238095238 |
| Financial Quantitative Analysts | 0.19047619 | 0.619047619 | 0.19047619 |
| Personal Financial Advisors | 0.19047619 | 0.666666667 | 0.142857143 |
| Retail Loss Prevention Specialists | 0.19047619 | 0.714285714 | 0.095238095 |
| Logisticians | 0.19047619 | 0.80952381 | 0 |
| Allergists and Immunologists | 0.1875 | 0.5625 | 0.25 |
| Anesthesiologist Assistants | 0.1875 | 0.5625 | 0.25 |
| Regulatory Affairs Specialists | 0.1875 | 0.59375 | 0.21875 |
| Electrical Drafters | 0.1875 | 0.6875 | 0.125 |

| Occupation | Col1 | Col2 | Col3 |
|---|---|---|---|
| Neurodiagnostic Technologists | 0.1875 | 0.6875 | 0.125 |
| Specialists | 0.1875 | 0.75 | 0.0625 |
| Pediatricians, General | 0.1875 | 0.75 | 0.0625 |
| Health Educators | 0.1875 | 0.8125 | 0 |
| Quality Control Systems Managers | 0.185185185 | 0.703703704 | 0.111111111 |
| Reporters and Correspondents | 0.185185185 | 0.777777778 | 0.037037037 |
| Except Special Education | 0.184210526 | 0.736842105 | 0.078947368 |
| Chiropractors | 0.181818182 | 0.272727273 | 0.545454545 |
| Mathematicians | 0.181818182 | 0.363636364 | 0.454545455 |
| Athletic Trainers | 0.181818182 | 0.590909091 | 0.227272727 |
| Equipment Installers and Repairers | 0.181818182 | 0.636363636 | 0.181818182 |
| Teachers, Secondary School | 0.181818182 | 0.727272727 | 0.090909091 |
| Funeral Service Managers | 0.181818182 | 0.727272727 | 0.090909091 |
| Podiatrists | 0.181818182 | 0.727272727 | 0.090909091 |
| Speech-Language Pathologists | 0.181818182 | 0.772727273 | 0.045454545 |
| Personal Care Aides | 0.181818182 | 0.818181818 | 0 |
| Prosthodontists | 0.181818182 | 0.818181818 | 0 |
| Treasurers and Controllers | 0.181818182 | 0.818181818 | 0 |
| Education and Literacy Teachers and Instructors | 0.179487179 | 0.743589744 | 0.076923077 |
| Surgical Assistants | 0.178571429 | 0.535714286 | 0.285714286 |
| Postsecondary | 0.178571429 | 0.571428571 | 0.25 |
| Preschool and Childcare Center/Program | 0.176470588 | 0.529411765 | 0.294117647 |
| Models | 0.176470588 | 0.529411765 | 0.294117647 |
| Technologists | 0.176470588 | 0.705882353 | 0.117647059 |
| Technicians | 0.176470588 | 0.705882353 | 0.117647059 |
| Sales Managers | 0.176470588 | 0.764705882 | 0.058823529 |
| Kindergarten and Elementary | 0.175 | 0.725 | 0.1 |
| Teachers, Postsecondary | 0.173913043 | 0.434782609 | 0.391304348 |
| Fitters | 0.173913043 | 0.695652174 | 0.130434783 |
| Millwrights | 0.173913043 | 0.782608696 | 0.043478261 |
| Program Directors | 0.173913043 | 0.782608696 | 0.043478261 |
| Petroleum Engineers | 0.173913043 | 0.826086957 | 0 |
| Technologists | 0.172413793 | 0.551724138 | 0.275862069 |
| Product Safety Engineers | 0.166666667 | 0.5 | 0.333333333 |
| Risk Management Specialists | 0.166666667 | 0.5 | 0.333333333 |
| Fire Inspectors | 0.166666667 | 0.583333333 | 0.25 |
| Airfield Operations Specialists | 0.166666667 | 0.666666667 | 0.166666667 |
| Compliance Managers | 0.166666667 | 0.666666667 | 0.166666667 |
| Mathematical Technicians | 0.166666667 | 0.666666667 | 0.166666667 |
| Nannies | 0.166666667 | 0.666666667 | 0.166666667 |

| Occupation | | | |
|---|---|---|---|
| Physician Assistants | 0.166666667 | 0.666666667 | 0.166666667 |
| Assistants | 0.166666667 | 0.666666667 | 0.166666667 |
| Chief Sustainability Officers | 0.166666667 | 0.722222222 | 0.111111111 |
| Technicians | 0.166666667 | 0.722222222 | 0.111111111 |
| Managers | 0.166666667 | 0.722222222 | 0.111111111 |
| Electrical Engineering Technicians | 0.166666667 | 0.75 | 0.083333333 |
| Repairers | 0.166666667 | 0.777777778 | 0.055555556 |
| Logistics Managers | 0.166666667 | 0.8 | 0.033333333 |
| Search Marketing Strategists | 0.166666667 | 0.805555556 | 0.027777778 |
| Anesthesiologists | 0.166666667 | 0.833333333 | 0 |
| Biomass Power Plant Managers | 0.166666667 | 0.833333333 | 0 |
| Database Administrators | 0.166666667 | 0.833333333 | 0 |
| and Investigators | 0.166666667 | 0.833333333 | 0 |
| Logistics Engineers | 0.166666667 | 0.833333333 | 0 |
| Political Scientists | 0.166666667 | 0.833333333 | 0 |
| Special Education | 0.162162162 | 0.810810811 | 0.027027027 |
| Biostatisticians | 0.16 | 0.72 | 0.12 |
| Inspectors | 0.157894737 | 0.631578947 | 0.210526316 |
| Specialists | 0.157894737 | 0.736842105 | 0.105263158 |
| Representatives and Officers | 0.157894737 | 0.736842105 | 0.105263158 |
| Investment Underwriters | 0.157894737 | 0.736842105 | 0.105263158 |
| School Psychologists | 0.157894737 | 0.736842105 | 0.105263158 |
| Tutors | 0.157894737 | 0.789473684 | 0.052631579 |
| Patient Representatives | 0.153846154 | 0.384615385 | 0.461538462 |
| Surgeons | 0.153846154 | 0.384615385 | 0.461538462 |
| Civil Engineering Technicians | 0.153846154 | 0.461538462 | 0.384615385 |
| Postsecondary | 0.153846154 | 0.538461538 | 0.307692308 |
| Sculptors, and Illustrators | 0.153846154 | 0.538461538 | 0.307692308 |
| Hospitalists | 0.153846154 | 0.538461538 | 0.307692308 |
| Cost Estimators | 0.153846154 | 0.615384615 | 0.230769231 |
| Workers | 0.153846154 | 0.615384615 | 0.230769231 |
| Insulation Workers, Mechanical | 0.153846154 | 0.615384615 | 0.230769231 |
| Avionics Technicians | 0.153846154 | 0.692307692 | 0.153846154 |
| Fire Investigators | 0.153846154 | 0.692307692 | 0.153846154 |
| Nuclear Medicine Physicians | 0.153846154 | 0.730769231 | 0.115384615 |
| Agents and Business Managers of Artists, Performers, and Athletes | 0.153846154 | 0.769230769 | 0.076923077 |
| Bailiffs | 0.153846154 | 0.769230769 | 0.076923077 |
| Mental Health Counselors | 0.153846154 | 0.769230769 | 0.076923077 |
| Photonics Engineers | 0.153846154 | 0.769230769 | 0.076923077 |
| Prevention Supervisors | 0.153846154 | 0.807692308 | 0.038461538 |
| Human Resources Managers | 0.153846154 | 0.807692308 | 0.038461538 |
| and Investigators | 0.153846154 | 0.846153846 | 0 |

| Occupation | | | |
|---|---|---|---|
| Massage Therapists | 0.153846154 | 0.846153846 | 0 |
| Solar Energy Systems Engineers | 0.153846154 | 0.846153846 | 0 |
| Structural Iron and Steel Workers | 0.15 | 0.55 | 0.3 |
| Conciliators | 0.15 | 0.65 | 0.2 |
| Dentists, General | 0.15 | 0.65 | 0.2 |
| Nuclear Engineers | 0.15 | 0.65 | 0.2 |
| Mechanics, Except Engines | 0.15 | 0.8 | 0.05 |
| Editors | 0.15 | 0.85 | 0 |
| Postsecondary | 0.15 | 0.85 | 0 |
| Postsecondary | 0.148148148 | 0.481481481 | 0.37037037 |
| Water/Wastewater Engineers | 0.148148148 | 0.851851852 | 0 |
| Astronomers | 0.142857143 | 0.571428571 | 0.285714286 |
| Molecular and Cellular Biologists | 0.142857143 | 0.571428571 | 0.285714286 |
| Cytotechnologists | 0.142857143 | 0.642857143 | 0.214285714 |
| Announcers | 0.142857143 | 0.642857143 | 0.214285714 |
| Actuaries | 0.142857143 | 0.714285714 | 0.142857143 |
| Microbiologists | 0.142857143 | 0.714285714 | 0.142857143 |
| Operations Research Analysts | 0.142857143 | 0.714285714 | 0.142857143 |
| Sustainability Specialists | 0.142857143 | 0.714285714 | 0.142857143 |
| Biochemical Engineers | 0.142857143 | 0.771428571 | 0.085714286 |
| Engineers/Architects | 0.142857143 | 0.857142857 | 0 |
| Marine Architects | 0.142857143 | 0.857142857 | 0 |
| Community Health Workers | 0.137931034 | 0.827586207 | 0.034482759 |
| Construction Carpenters | 0.136363636 | 0.590909091 | 0.272727273 |
| Instructors | 0.136363636 | 0.590909091 | 0.272727273 |
| Biologists | 0.136363636 | 0.681818182 | 0.181818182 |
| Validation Engineers | 0.136363636 | 0.818181818 | 0.045454545 |
| Recreation Workers | 0.136363636 | 0.863636364 | 0 |
| Curators | 0.133333333 | 0.466666667 | 0.4 |
| Security Managers | 0.133333333 | 0.566666667 | 0.3 |
| Epidemiologists | 0.133333333 | 0.733333333 | 0.133333333 |
| Legislators | 0.133333333 | 0.8 | 0.066666667 |
| Supply Chain Managers | 0.133333333 | 0.8 | 0.066666667 |
| Workers | 0.130434783 | 0.739130435 | 0.130434783 |
| Dietitians and Nutritionists | 0.130434783 | 0.782608696 | 0.086956522 |
| Sales Agents, Financial Services | 0.125 | 0.125 | 0.75 |
| Managers | 0.125 | 0.583333333 | 0.291666667 |
| Atmospheric and Space Scientists | 0.125 | 0.666666667 | 0.208333333 |
| Art Directors | 0.125 | 0.75 | 0.125 |
| Analysis Specialists | 0.125 | 0.791666667 | 0.083333333 |
| Healthcare Social Workers | 0.125 | 0.8125 | 0.0625 |
| Green Marketers | 0.125 | 0.875 | 0 |
| Rehabilitation Counselors | 0.125 | 0.875 | 0 |

| Occupation | | | |
|---|---|---|---|
| Computer Network Architects | 0.121212121 | 0.757575758 | 0.121212121 |
| Automotive Engineers | 0.12 | 0.84 | 0.04 |
| Foresters | 0.12 | 0.84 | 0.04 |
| Adjusters | 0.117647059 | 0.529411765 | 0.352941176 |
| Stonemasons | 0.117647059 | 0.882352941 | 0 |
| Transportation Engineers | 0.115384615 | 0.769230769 | 0.115384615 |
| Quality Control Analysts | 0.115384615 | 0.846153846 | 0.038461538 |
| Ergonomists | 0.115384615 | 0.884615385 | 0 |
| Athletes and Sports Competitors | 0.111111111 | 0 | 0.888888889 |
| Technicians | 0.111111111 | 0.666666667 | 0.222222222 |
| Ophthalmologists | 0.111111111 | 0.666666667 | 0.222222222 |
| Nurse Practitioners | 0.111111111 | 0.703703704 | 0.185185185 |
| Sports Medicine Physicians | 0.111111111 | 0.851851852 | 0.037037037 |
| Anthropologists | 0.107142857 | 0.607142857 | 0.285714286 |
| Transportation Managers | 0.107142857 | 0.714285714 | 0.178571429 |
| Labor Relations Specialists | 0.107142857 | 0.75 | 0.142857143 |
| Postsecondary | 0.107142857 | 0.892857143 | 0 |
| Dermatologists | 0.105263158 | 0.473684211 | 0.421052632 |
| Managers | 0.105263158 | 0.736842105 | 0.157894737 |
| Managers | 0.105263158 | 0.736842105 | 0.157894737 |
| Respiratory Therapy Technicians | 0.105263158 | 0.789473684 | 0.105263158 |
| Urban and Regional Planners | 0.105263158 | 0.789473684 | 0.105263158 |
| Genetic Counselors | 0.105263158 | 0.842105263 | 0.052631579 |
| Optometrists | 0.1 | 0.5 | 0.4 |
| Bioinformatics Scientists | 0.1 | 0.65 | 0.25 |
| Music Therapists | 0.1 | 0.8 | 0.1 |
| Specialists | 0.1 | 0.8 | 0.1 |
| and Mobility Specialists, and Vision Rehabilitation Therapists | 0.1 | 0.9 | 0 |
| Historians | 0.095238095 | 0.523809524 | 0.380952381 |
| Water Resource Specialists | 0.095238095 | 0.666666667 | 0.238095238 |
| Computer Systems Analysts | 0.095238095 | 0.761904762 | 0.142857143 |
| Clinical Psychologists | 0.095238095 | 0.904761905 | 0 |
| Engineers | 0.095238095 | 0.904761905 | 0 |
| Correctional Officers | 0.090909091 | 0.681818182 | 0.227272727 |
| Agricultural Inspectors | 0.090909091 | 0.909090909 | 0 |
| Interior Designers | 0.090909091 | 0.909090909 | 0 |
| and Vocational Counselors | 0.088235294 | 0.764705882 | 0.147058824 |
| Biochemists and Biophysicists | 0.086956522 | 0.782608696 | 0.130434783 |
| Producers | 0.083333333 | 0.375 | 0.541666667 |
| Fish and Game Wardens | 0.083333333 | 0.625 | 0.291666667 |
| Dietetic Technicians | 0.083333333 | 0.666666667 | 0.25 |
| Economists | 0.083333333 | 0.75 | 0.166666667 |

| Occupation | | | |
|---|---|---|---|
| Psychiatrists | 0.083333333 | 0.75 | 0.166666667 |
| Video Game Designers | 0.083333333 | 0.75 | 0.166666667 |
| and Technologists | 0.083333333 | 0.833333333 | 0.083333333 |
| Motorcycle Mechanics | 0.083333333 | 0.833333333 | 0.083333333 |
| Segmental Pavers | 0.083333333 | 0.833333333 | 0.083333333 |
| Naval | 0.083333333 | 0.916666667 | 0 |
| Chemical Engineers | 0.083333333 | 0.916666667 | 0 |
| Managers | 0.083333333 | 0.916666667 | 0 |
| Marketing Specialists | 0.076923077 | 0.461538462 | 0.461538462 |
| Fuel Cell Engineers | 0.076923077 | 0.769230769 | 0.153846154 |
| Budget Analysts | 0.076923077 | 0.846153846 | 0.076923077 |
| Counseling Psychologists | 0.076923077 | 0.846153846 | 0.076923077 |
| Adjudicators, and Hearing Officers | 0.076923077 | 0.923076923 | 0 |
| Engineers | 0.076923077 | 0.923076923 | 0 |
| Abuse Social Workers | 0.076923077 | 0.923076923 | 0 |
| Talent Directors | 0.076923077 | 0.923076923 | 0 |
| Set and Exhibit Designers | 0.074074074 | 0.888888889 | 0.037037037 |
| Oral and Maxillofacial Surgeons | 0.071428571 | 0.857142857 | 0.071428571 |
| Advisors | 0.066666667 | 0.533333333 | 0.4 |
| Research Scientists | 0.066666667 | 0.733333333 | 0.2 |
| Musicians, Instrumental | 0.066666667 | 0.733333333 | 0.2 |
| Preventive Medicine Physicians | 0.066666667 | 0.8 | 0.133333333 |
| Obstetricians and Gynecologists | 0.066666667 | 0.866666667 | 0.066666667 |
| Rehabilitation Physicians | 0.066666667 | 0.933333333 | 0 |
| Microsystems Engineers | 0.064516129 | 0.677419355 | 0.258064516 |
| Orthoptists | 0.0625 | 0.4375 | 0.5 |
| Psychiatric Technicians | 0.0625 | 0.5 | 0.4375 |
| Wind Energy Engineers | 0.0625 | 0.75 | 0.1875 |
| Managers | 0.0625 | 0.8125 | 0.125 |
| Fuel Cell Technicians | 0.0625 | 0.9375 | 0 |
| Marriage and Family Therapists | 0.0625 | 0.9375 | 0 |
| Geothermal Production Managers | 0.058823529 | 0.647058824 | 0.294117647 |
| Informatics Nurse Specialists | 0.058823529 | 0.647058824 | 0.294117647 |
| Choreographers | 0.055555556 | 0.388888889 | 0.555555556 |
| Public Relations Specialists | 0.055555556 | 0.722222222 | 0.222222222 |
| Music Directors | 0.055555556 | 0.777777778 | 0.166666667 |
| Neuropsychologists | 0.055555556 | 0.833333333 | 0.111111111 |
| Aerospace Engineers | 0.055555556 | 0.944444444 | 0 |
| Education | 0.052631579 | 0.421052632 | 0.526315789 |
| Environmental Economists | 0.052631579 | 0.736842105 | 0.210526316 |
| Instructional Coordinators | 0.052631579 | 0.842105263 | 0.105263158 |
| Landscape Architects | 0.052631579 | 0.947368421 | 0 |
| Technologists | 0.05 | 0.9 | 0.05 |

| Occupation | Col1 | Col2 | Col3 |
|---|---|---|---|
| Marketing Managers | 0.05 | 0.9 | 0.05 |
| Correctional Treatment Specialists | 0.047619048 | 0.571428571 | 0.380952381 |
| Managers | 0.047619048 | 0.80952381 | 0.142857143 |
| Electrical Engineers | 0.045454545 | 0.681818182 | 0.272727273 |
| Specialists and Site Managers | 0.045454545 | 0.954545455 | 0 |
| Specialists, Including Health | 0.045454545 | 0.954545455 | 0 |
| Disorder Counselors | 0.043478261 | 0.52173913 | 0.434782609 |
| Neurologists | 0.041666667 | 0.625 | 0.333333333 |
| Physical Therapists | 0.041666667 | 0.708333333 | 0.25 |
| Television, and Radio | 0.04 | 0.72 | 0.24 |
| Psychologists | 0.04 | 0.84 | 0.12 |
| Soil and Plant Scientists | 0.037037037 | 0.962962963 | 0 |
| Coaches and Scouts | 0.035714286 | 0.785714286 | 0.178571429 |
| Clinical Nurse Specialists | 0.034482759 | 0.896551724 | 0.068965517 |
| Chief Executives | 0.032258065 | 0.612903226 | 0.35483871 |
| Elementary and Secondary School | 0.032258065 | 0.709677419 | 0.258064516 |
| and Geographers | 0.032258065 | 0.935483871 | 0.032258065 |
| Clergy | 0 | 0.428571429 | 0.571428571 |
| Dancers | 0 | 0.538461538 | 0.461538462 |
| Sociologists | 0 | 0.583333333 | 0.416666667 |
| Designers | 0 | 0.588235294 | 0.411764706 |
| Singers | 0 | 0.615384615 | 0.384615385 |
| Writers | 0 | 0.625 | 0.375 |
| Managers | 0 | 0.714285714 | 0.285714286 |
| Directors | 0 | 0.739130435 | 0.260869565 |
| Lawyers | 0 | 0.772727273 | 0.227272727 |
| Supervisors | 0 | 0.8 | 0.2 |
| Magistrates | 0 | 0.8 | 0.2 |
| Nanosystems Engineers | 0 | 0.8 | 0.2 |
| Orthodontists | 0 | 0.8 | 0.2 |
| Natural Sciences Managers | 0 | 0.8125 | 0.1875 |
| Range Managers | 0 | 0.875 | 0.125 |
| Animal Scientists | 0 | 0.888888889 | 0.111111111 |
| Biofuels/Biodiesel Technology and Product Development Managers | 0 | 0.947368421 | 0.052631579 |
| Climate Change Analysts | 0 | 1 | 0 |
| Copy Writers | 0 | 1 | 0 |
| Data Warehousing Specialists | 0 | 1 | 0 |
| Planners | 0 | 1 | 0 |
| Industrial Ecologists | 0 | 1 | 0 |
| Technologists | 0 | 1 | 0 |
| Recreational Therapists | 0 | 1 | 0 |
| Regulatory Affairs Managers | 0 | 1 | 0 |

| Wind Energy Project Managers | 0 | 1 | 0 |

APPENDIX B

THE RESULTS OF INDUSTRIES' AUTOMATABILITY

| Industries | Substitution | Complementarity | Negligibility |
|---|---|---|---|
| Educational Services | 0.230367742 | 0.543841067 | 0.225791191 |
| Arts, Entertainment and Recreation | 0.273589262 | 0.534721882 | 0.191688856 |
| Other Services (Except Public Administration) | 0.334531855 | 0.522794247 | 0.142673898 |
| Real Estate and Rental and Leasing | 0.332822567 | 0.533638338 | 0.133539095 |
| Health Care and Social Assistance | 0.282281819 | 0.590504851 | 0.12721333 |
| Government | 0.284248443 | 0.60369429 | 0.112057267 |
| Professional, Scientific and Technical Services | 0.239996052 | 0.650227068 | 0.10977688 |
| Agriculture, Forestry, Fishing and Hunting | 0.415514488 | 0.4798687 | 0.104616812 |
| Construction | 0.402369766 | 0.499573624 | 0.09805661 |
| Management of Companies and Enterprises | 0.219898769 | 0.688056181 | 0.09204505 |
| Information | 0.337971671 | 0.575005684 | 0.087022645 |
| Manufacturing | 0.447618473 | 0.467541898 | 0.084839629 |
| Utilities | 0.34821794 | 0.570123881 | 0.081658179 |
| Mining, Quarrying, and Oil and Gas Extraction | 0.466154889 | 0.452643383 | 0.081201727 |
| Finance and Insurance | 0.369701562 | 0.554126927 | 0.07617151 |
| Administrative and Support Services | 0.544294248 | 0.386182565 | 0.069523187 |
| Wholesale Trade | 0.402604274 | 0.535910322 | 0.061485404 |
| Retail Trade | 0.479448168 | 0.459071623 | 0.061480209 |
| Transportation and Warehousing | 0.442244448 | 0.49711725 | 0.060638302 |
| Accommodation and Food Services | 0.597011386 | 0.348796044 | 0.05419257 |


\end{appendices}



\end{document}